\documentclass[reqno]{amsart}
\usepackage{amsmath}
\usepackage{amsthm}
\usepackage{amssymb}
\usepackage[
left=1.25in, right=1.25in]{geometry}
\usepackage{tikz}
\usepackage[all]{xy}
\usepackage{graphicx}
\usepackage{mathrsfs}
\usepackage{hyperref}
\usepackage{microtype}
\usepackage{url,doi}
\usepackage{braket}
\usepackage{ifpdf}
\usepackage{mathtools}
\usepackage{color,soul}

\usepackage[figuresright]{rotating}
\usepackage{bm}
\include{epsf}
\usepackage{color}
\usepackage{xcolor}
\usepackage{braket}
\usepackage{amsmath,amssymb,amsthm,mathrsfs,amsfonts,dsfont,bm,mathtools}  
\usepackage{graphicx}
\usepackage{subfigure}
\usepackage[titletoc]{appendix}
\DeclareSymbolFont{symbolsC}{U}{pxsyc}{m}{n}
\DeclareMathSymbol{\coloneqq}{\mathrel}{symbolsC}{"42}

\newcommand{\Z}{{\mathscr Z}}
\newcommand{\F}{{\mathscr F}}

\newcounter{mycite}
\newtoks\citetoks
\makeatletter
\DeclareRobustCommand\unscite[1]{%
  \@ifundefined{uns@cite#1}
    {\refstepcounter{mycite}\label{citelabel@#1}%
     \expandafter\xdef\csname uns@cite#1\endcsname{\arabic{mycite}}%
     \toks\z@=\expandafter{\the\citetoks}%
     \toks\tw@=\expandafter\expandafter\expandafter{%
       \csname uns@bibitem#1\endcsname}%
     \edef\@tempcite{\the\toks\z@\the\toks\tw@}%
     \global\citetoks=\expandafter{\@tempcite}%
    }{}[\@nameuse{uns@cite#1}]}
\newcommand{\mybibitem}[2]{%
  \@namedef{uns@bibitem#1}{\bibitem[\ref{citelabel@#1}]{#1}#2}}
\makeatother
\usepackage{verbatim}
\usepackage{graphicx,adjustbox}

\begin{document}

\title{Phase Structure of XX0 Spin Chain and Nonintersecting Brownian Motion}

\author{M. Saeedian}
\address{Department of Physics, Shahid Beheshti University, G.C., Evin, Tehran 19839, Iran}

\author{A. Zahabi}
\address{Department of Physics, Witwatersrand University, Johannesburg, South Africa}


\begin{abstract}
We study finite size and temperature XX0 Heisenberg spin chain in weak and strong coupling regimes. By using an elegant connection of the model to integrable combinatorics and probability, we explore and interpret a possible phase structure of the model in asymptotic limit: the limit of large inverse temperature and size. First, the partition function and free energy of the model are derived by using techniques and results from random matrix models and nonintersecting Brownian motion. We show that, in the asymptotic limit, partition function of the model, written in terms of matrix integral, is governed by the Tracy-Widom distribution. Second, the exact analytic results for the free energy, which is obtained by the asymptotic analysis of the Tracy-Widom distribution, indicate a completely new and sophisticated phase structure of the model. This phase structure consists of second- and third-order phase transitions. Finally, to shed light on our new results, we provide a possible new interpretation of the phase structure in terms of dynamical behaviour of magnons in the spin chain. We demonstrate distinct features of the phases with schematic spin configurations which have definite features in each region of the phase diagram.
\end{abstract}

\maketitle

\section{Introduction}
\label{s1}
The Heisenberg spin chain is a one-dimensional lattice model consisting of (half) integer spins interacting with their neighboring spins and an external magnetic field \unscite{baxter}. This model, being among the original models for studying the quantum magnetic properties of the matter, has been studied during this and last century, extensively, for a review see \unscite{Schollwock}. Moreover, the Heisenberg model is one of the pillars of quantum integrability and is among the most famous exactly solvable lattice models in quantum statistical mechanics. In fact, this model is a theoretical laboratory for testing and applying the integrability paradigm. The exact methods and techniques from quantum statistical mechanics and quantum field theory such as free fermion representation \unscite{Lieb}, bosonization \unscite{schulz}, and Bethe ansatz \unscite{faddeev} have been applied in Heisenberg model, extensively.

The Heisenberg model is a general name for a class of spin chains which are characterized by three couplings $\{J_x, J_y, J_z\}$, associated to three components of the spin operators in the Hamiltonain of the system, Eq. \ref{H_XYZ}. According to the couplings, the most general Heisenberg model with $J_x\neq J_y\neq J_z$ is called XYZ model and special cases include XXZ, XY models, etc. Due to complexity of the abstract space of parameters ($\{J_x, J_y, J_z, h\}$), these spin chains exhibit sophisticated phase structure and rich phenomena, including thermal, topological and quantum phase transitions \unscite{Schollwock},\unscite{sachdev}. Moreover, the spin chains are applied in variety of different systems, from condensed matter physics to high energy physics, such as quantum information theory \unscite{eisert}, and (non)supersymmetric gauge theories and AdS/CFT \unscite{minahan},\unscite{nekrasov1},\unscite{nekrasov2}.

The subject of this study is the simplest case of the Heisenberg model, namely the XX0 model, for the definition see Eq. \ref{H_XX0 eq1} and the explanations around that. Although this model is simple, it is of great interest from different perspectives. In fact, this model has interesting and nontrivial features such as phase transitions. On the other hand, the simplicity of the model leads to its exact solvability, via the free fermion formalism. This formalism, using the Jordan-Wigner transformation, makes the model as one of the basic important examples in the class of integrable systems. 

From a different point of view, the XX0 model is closely related to integrable combinatorics and probability, i.e. the class of integrable combinatorial lattice models \unscite{bogolyobovRev}, \unscite{borodin}, \unscite{di}. These are two- and three-dimensional integrable lattice models such as dimer models, nonintersecting Brownian motion and plane partitions (crystal melting models), etc. Furthermore, the mathematical structure of the XX0 model reveals an exact and concrete relation to the random matrix theory \unscite{blower}, \unscite{forrester}. In this study, the above mentioned relations between XX0 model, nonintersecting Brownian motion and random matrix theory provide powerful techniques to investigate different aspects of the XX0 model. 

The random matrix theory is applied in statistical physics and quantum field theories. In fact, the statistical lattice models such as spin chains are very suitable for implementing the methods of random matrix models and their discrete versions. Especially, the asymptotic limits of the statistical systems such as spin chains can be accurately described by the asymptotic analysis of the associated matrix models. In this article, the explicit realization of the XX0 model in terms of the matrix integral with the Gross-Witten potential \unscite{Bogoliubov},\unscite{Bogoliubov1},\unscite{Bogoliubov2}, is one the building blocks of our study.

The nonintersecting Brownian motion is one-dimensional random walk subject to a constraint, i.e. the paths of random walkers do not intersect. This model has been studied extensively for decades and various exact mathematical and physical results have been obtained \unscite{TW1}, \unscite{fisher}, \unscite{krattenthaler}, \unscite{baik}. The extensive studies and results in nonintersecting Brownian motion have been established this subject in the new trend of integrable probability. This integrable lattice model exhibits different aspects and features in different mathematical and physical subjects such as combinatorics, probability and integrable statistical systems. Recently, it has been shown by Bogoliubov that the path configurations of the nonintersecting Brownian motion are in one-to-one correspondence with the spin configurations of XX0 spin chain. Furthermore, recent studies in probabilistic and asymptotic aspects of nonintersecting Brownian motion \unscite{Baik}, \unscite{Majumdar}, \unscite{Schehr1}, \unscite{Forrester}, and \unscite{Kobayashi} reveal the integrability and universality structures in this model. They are originated from the appearance of Toeplitz determinants and Tracy-Widom distribution.

Random matrix theory from its first appearance in 1930's, has been used in different branches of mathematics and physics. In mathematical physics, it has been mainly employed in statistical mechanics and quantum field theories to calculate the correlation functions of systems with complicated interactions. These techniques have been expanded, refined and enriched in various different problems from high energy physics such as Yang-Mill theory \unscite{RM QCD}, and quantum gravity \unscite{RM Gravity} to statistical physics such as diffusion process \unscite{RM nuclear}, NIBM \unscite{baik}, traffic and communication network \unscite{Rm network}, and stock movements in financial markets \unscite{RM finance1},\unscite{RM finance5}, etc. In this section we will apply this technique to tackle the correlation functions of XX0 model.

Applying the machinery of the random matrix models and nonintersecting Brownian motion, we study the XX0 model in weak and strong coupling regimes. Using these methods, we obtain our new results. More precisely, we explicitly calculate the partition function and free energy of the model and we extract phase structure of the models, in asymptotic regime. Based on obtained mathematical results, appropriate and plausible interpretations for the phase structure of XX0 model are discussed.

At the heart of this study, we use the analytic and probabilistic methods from random matrix models and nonintersecting Brownian motion to obtain the exact new results in XX0 model and explain the physical features of the model. In the first step, we review a representation for the partition functions of the infinite/finite XX0 model in terms of the matrix integrals and corresponding Toeplitz determinants, and their discrete versions. We obtain that the partition functions of strongly and weakly coupled XX0 models are given by matrix integrals with Gross-Witten and quadratic potentials, respectively.

The main technique for studying the free energy and phase structure of the finite XX0 model is extracted from the recent results in the context of the nonintersecting Brownian motion. The exact relations between the nonintersecting Brownian motion and XX0 in \unscite{Bogoliubov},\unscite{Bogoliubov1},\unscite{Bogoliubov2}, pave the way for applying new probabilistic methods and results \unscite{Baik} from nonintersecting Brownian motion to XX0 model. This leads to new results in the XX0 model.
In fact, we use these methods to study and interpret the previously obtained results for free energy and phase structure of the infinite XX0 model \unscite{perez}. Moreover, we perform a completely new study about the finite XX0 model, its finite size effects in evaluation of free energy, phase structure and their interpretations. We exactly follow this approach and obtain exact analytic results for the free energy and phase structure of the finite XX0 model in the asymptotic limit. More precisely, we use the obtained results from the definite stochastic processes in nonintersecting Brownian motion, which are expressed in terms of the continuous/discrete Toeplitz determinants, and then apply them directly to calculate the free energy of XX0 model. 

Following this approach, we obtain our new results indicating that in the asymptotic limit, ratio of the partition functions of finite and infinite XX0 spin chain is given by the Tracy-Widom distribution. Furthermore, we elaborate on asymptotic limit of finite XX0 by precise analysis of the Tracy-Widom distribution in the asymptotic limit. We find a new, rich and sophisticated phase structure that exhibits different physical features in different regions of the moduli space of the parameters. More precisely, we separately study the asymptotic limit of finite XX0 model in strong and weak coupling regimes and we obtain explicit new expressions for free energy in each regime. The obtained results indicate: (I) a phase diagram with second- and third-order domain walls in strong coupling regime and (II) a third-order phase transition in weak coupling regime.

Finally, we use the basic connection between XX0 model and nonintersecting Brownian motion to obtain a mathematical result for our interpretation. Based on this connections and the obtained mathematical result, we provide a possible new interpretation for free energy and phase structure of the XX0 model. This interpretation is in terms of the dynamics of the magnons and their distribution in the spin chain. This distribution determines a quantum state which is a representative of each region in the phase diagram. Via this, first we interpret the Gross-Witten phase transition in the infinite XX0 model by the phenomenon of freezing of the magnons in the spin chain. Then, we interpret the extra term in the free energy of the finite XX0 model as the finite size effect and by using the mathematical results about the dynamics of the diffusing magnons we find interpretation of the finite size effect. 

This paper is organized as follows. In sec. II, we describe finite XX0 model and its matrix model representation. New results in free energy and phase structure of finite XX0 model in the asymptotic limit are presented in sec. III. 
Then, in sec. IV, we interpret the obtained results about the new phase transitions in XX0 model.
Finally, in sec. V we discuss possible issues and available directions for further studies. To be self-contained, there are four appendices that cover some physical and mathematical results from the literature that are crucial for understanding the new results of the paper. 

\section{XX0 Heisenberg spin chain and random matrix theory}
\label{s2}
In this section we define the Heisenberg XX0 model and briefly review some of results without derivation. Our main focus is on the partition function of the model and its representation in terms of the matrix integrals and Toeplitz determinants. However, as it will be crucial in the rest of this paper, we will point out some similarities between dynamics of the XX0 model and that of nonintersecting Brownian motion (NIBM). We keep our discussions in this section to a minimum required length and thus we only discuss the necessary ideas and results that will be used in the following chapters. To be self-contained, we collect some crucial facts and results in Appendices and for further explanations and derivations of the results, see for example \unscite{bogolyobovRev}.

The XX0 model has diverse relations with recent topics of research in mathematical physics such as combinatorial and probabilistic models such as alternating sign matrices \unscite{bressoud}, random tilings, theory of random walks in lattice and random matrix theory \unscite{baik}, plane partitions and theory of symmetric functions \unscite{macmahon},\unscite{macdonald}, and also topological string theory \unscite{okounkov}.

\subsection{XX0 spin chain and its correlation functions}
Let us start with the most general finite size, XYZ Heisenberg spin chain with the following Hamiltonian,
\begin{eqnarray}
\label{H_XYZ}\nonumber
{\hat{H}}_{XYZ}=-\sum_{k=0}^{N}\bigg(J_{x}{\sigma }_{k}^{x}{\sigma }_{k+1}^{x}+J_{y}{\sigma }_{k}^{y}{\sigma }_{k+1}^{y}+J_{z}{\sigma }_{k}^{z}{\sigma }_{k+1}^{z}+h{\sigma }_{k}^{z}\bigg),
\end{eqnarray}
where ${\sigma}^{x,y,z}$ are the Pauli matrices,
\[
\sigma^{x}=
\begin{bmatrix}
    0       & 1   \\
    1       & 0   \\
\end{bmatrix},
\quad
\sigma^{y}=
\begin{bmatrix}
    0       & -i   \\
    i       & 0   \\
\end{bmatrix},
\quad
\sigma^{z}=
\begin{bmatrix}
    1       & 0   \\
    0       & -1   \\
\end{bmatrix},
\]
$J_x, J_y, J_z$ are the couplings and $h$ is the external magnetic field.
In this study we focus on the isotropic and periodic Heisenberg spin chain of finite size $N$ with $J_x=J_y=\Delta/8$, $J_z=0$ and $h=0$, which is called XX0 spin chain with the following Hamiltonian,
\begin{eqnarray}
\label{H_XX0 eq1}
{\hat{H}}_{XX0}=-\sum_{n=1}^{N}\sum_{m=1}^N{\Delta_{nm}}{\sigma }_{n}^{+}{\sigma }_{m}^{-},
\end{eqnarray}
where raising and lowering spin operators are defined by ${\sigma}_k^{\pm}=({\sigma}_k^{x}{\pm}i{\sigma}_k^{y})/2$, and ${\Delta_{nm}}$ is the nearest neighbour coupling constant, satisfying the periodic boundary condition,
\begin{eqnarray}
\label{Delta eq2}
{\Delta_{nm}}=\frac{\Delta}{2}({\delta_{|n-m|,1}}+{\delta_{|n-m|, N-1}}).
\end{eqnarray}

Recently, considerable studies have been devoted to different aspects of the XX0 model as an integrable model. This includes the study of time-dependent (thermal) correlation functions of XX0 model. The correlation functions of XX0 model as a special case of XY model, at zero temperature are studied in \unscite{Vaidya},\unscite{McCoy},\unscite{McCoy1}.
At nonzero temperature, by using the similarities to the gas of bosons on a line, the correlation functions of XX0 model are investigated in \unscite{Its},\unscite{Its1},\unscite{Its2},\unscite{Its3},\unscite{Its4},\unscite{Izergin}, and the equal time correlation function was obtained in \unscite{Lenard0},\unscite{Lenard}. Furthermore, time-dependent correlation function is obtained in \unscite{Korepin} using the Fredholm determinant of a linear integral operator. The correlation functions of XXZ and XX0 models in thermodynamic limit are studied in \unscite{kitanine},\unscite{colomo1},\unscite{colomo2},\unscite{colomo3},\unscite{Korepin}. Using similarities between XX0 model and NIBM, correlation functions of XX0 model in low temperature limit are recently formulated and investigated by Bogoliubov, \unscite{Bogoliubov},\unscite{Bogoliubov1},\unscite{Bogoliubov2}.

As we mentioned before and also summarized in Appendix D, different nonzero thermal correlation functions in the XX0 spin chain can be defined and studied, using the combinatorial features of the XX0 model. These correlation functions are basically made by spin averages over the product of spin operators acting on the ferromagnetic vacuum. We can rewrite the correlation functions by first acting the spin operators on the ferromagnetic vacuum and obtaining a new states and then the correlation functions can be written as the spin average over the time evolution operator between the new states.
In this study, first we fix the initial and final spin state. In other words, we consider a specific arrangement of the spin operators which leave us with a particular state, a set of $N_{f}$ neighboring spin-down states, located on a finite segment of ferromagnetic vacuum with length \textit{N}. This defines a correlation function, which for reasons that will be mentioned we call it the partition function from now on. Thus, the partition function is defined as
\begin{eqnarray}
\label{Delta eq7}
\Z_{XX0}\coloneqq\bra{\Uparrow}\prod_{i=1}^{N_f}\sigma^{+}_{N_f-i}e^{-t{\hat{H}}_{XX0}}\prod_{i=1}^{N_f}\sigma^{-}_{N_f-i}\ket{\Uparrow}=\bra{\overbrace{{\uparrow},...,{\uparrow},\underbrace{{\downarrow},...,{\downarrow},{\downarrow}}_{N_f}}^{N}}e^{-t{\hat{H}}_{XX0}}\ket{\overbrace{\underbrace{{\downarrow},{\downarrow},...,{\downarrow}}_{N_f},{\uparrow},...,{\uparrow}}^{N}},
\end{eqnarray}
where the ferromagnetic vacuum, the state of all the spins up, is defined as
$\ket{\Uparrow}={\otimes}_{n=1}^{N}\ket{\uparrow}_{n}$, and \textit{N}, \textit{$N_{f}$} and \textit{t} are  size of the spin chain, size of spin-down segment i.e. number of the flipped spins (in the literature, \unscite{Schollwock}, this is often called $N_f$ $\textit{magnons}$) and evolutionary parameter or time, respectively. It is natural to interpret the time as inverse temperature in the context of the statistical mechanics.

There are technical reasons for choosing this special correlation function. This correlation function is basically the simplest possible correlation function from the mathematical point of view. In fact, as we discussed in Appendix D, with this arrangement of spin operators, the Schur functiona, which appear in the matrix integral form of the correlation function, are equal to identity and thus the correlation function takes the simple form of the Gross-Witten matrix integral. Taking this into account and also considering the fact that $\hat{H}_{XX0}\ket{\Uparrow}=0$, and $\bra{\Uparrow}\exp(-t\hat{H}_{XX0})\ket{\Uparrow}=1$, we call the correlation function \ref{Delta eq7}, the partition function of XX0 model. From physical point of view, the prepared state with above arrangement of spin-downs has maximum energy in comparison with other states with equal number of magnons. The time evolution operator acts on this state and push the magnons to diffuse into the vacuum and reduce the energy. As we will explain in sec. 4, there will be different situations regarding to the distribution of magnons and this determines the phase structure of the model.

As we mentioned, the correlation functions of XX0 in the thermodynamic limit have been studied before. Our main goal in this study is to find the analytic expression for the above partition function (\ref{Delta eq7}) in the asymptotic limit; the limit of large $t$ , $N$ and  $N_f$. As we will observe, the result of asymptotic analysis of the partition function depends explicitly on the relations between the parameters or technically the moduli space of the parameters.

Our strategy toward the calculations of above partition function is first to use the matrix integral representation of the correlation functions and second to apply the results from NIBM, which is obtained in terms of the matrix integrals and Toeplitz determinants.

The relation between XX0 model and NIBM is studied comprehensively by Bogoliubov \unscite{Bogoliubov} and it has been used to derive the thermal correlation functions. This relation is at the heart of this study and thus let us summarize the key points of this connection. A remarkable property of the XX0 model is that the Hamiltonian of the model generates the dynamics of the one-dimensional vicious random walks. In fact, the transition between up and down spin at the sites of the XX0 spin chain can be considered as a random move of random walkers in an one-dimensional vicious random walks model, see Appendix A. Moreover, generating function for the vicious random walks can be constructed from the correlation functions of the spin operators in the ferromagnetic vacuum state of the XX0 model in a zero magnetic field.  Another remarkable feature of the XX0 model is that the the correlations functions, made out of the many particle states, have definite combinatorial features and in fact can be written as matrix integrals with Schur symmetric functions playing the role of spin operators. These properties and their consequences provide a feasible approach toward first, the understanding of a simple combinatorial and stochastic picture for the dynamics of the XX0 model and second, an explicit calculation of the correlation functions in the regime of finite parameters and also in the asymptotic limit.
\subsection{Matrix model for XX0 spin chain} \label{s3}

In this part, we explain the matrix integral and Toeplitz/Hankel determinant representation of the partition function of finite and infinite XX0 spin chains.
In this way, first, following \unscite{Bogoliubov}, we explain a result, indicating that the partition function of infinite XX0 spin chain can be represented by matrix integral with an specific weight function. Then we deduce the representation of the finite XX0 model as a discrete version of the matrix integral. More explanations on this matter are reviewed in Appendix D.
Finally, by using Heine-Szeg\"{o} identity \unscite{Heine} we express the partition function of XX0 model in terms of the Toeplitz and Hankel determinants.

The random matrix theory, the theory of finite/infinite matrices with Gaussian independently distributed random entries, is a powerful method for studying variety of physical systems and mathematical models. The main theme of applications of the random matrix theory in physics is emerged from the fact that the partition function of a two-dimensional gas of charged particles, interacting with a two dimensional Coulomb force and an external potential, can be extracted by a matrix integral, i.e. integration over all the eigenvalues of a random matrix in random Gaussian ensembles \unscite{Mehta org}. In the context of our study, the random matrix theory is used to extract the partition function of the XX0 spin chain. Using the connections with NIBM, Bogoliubov \unscite{Bogoliubov} derived the time dependent (thermal) correlation functions of the infinitely large size XX0 model, which we call it infinite XX0 model, in terms of a continuous matrix integral with Gross-Witten potential over the Schur functions, see Appendix \ref{PfCf}. As it has been argued above and in Appendix \ref{PfCf}, the partition function of the infinite XX0 model, Eq. \ref{Delta eq7}, for $N\rightarrow \infty$, is an special case (with the Schur functions equal to identity) of the matrix integral formula Eq. \ref{CM} for the correlation functions Eq. \ref{C},
\begin{eqnarray}
\label{matrix continous}
\Z_{XX0}=\prod_{j=1}^{N_{f}} \int_{-\pi}^{\pi} \frac{d\alpha_{j}}{2\pi} f_{GW}(e^{i\alpha_{j}}) \prod_{l<p} | e^{i\alpha_{l}}-e^{i\alpha_{p}}|^{2},\quad f_{GW}=e^{tV_{GW}},
\end{eqnarray}
where $f_{GW}$ is the weight function of the Gross-Witten potential $V_{GW}(z)=\frac{z+z^{-1}}{2}$ and $z=e^{i\alpha}$ ($\alpha$ is eigenvalue value of random matrix). Notice that this matrix model has dimension $N_{f}$ and \textit{t} is the parameter in the weight function. This matrix model has been studied in two-dimensional lattice gauge theories, called Gross-Witten-Wadia ( for short GW) model, \unscite{Wadia}, \unscite{Gross}. From now on, for more convenience we rename $\Z_{XX0}$ by $\Z_{GW}$.

It would be natural to think that the partition function of finite size XX0 model can be written as a discretization of the above matrix integral (\ref{matrix continous}) as
\begin{eqnarray}
\label{matrix discrete}
\Z_{GW}^{\mid d \mid}=\frac{1}{N_{f}!|d|^{N_{f}}} \sum_{(z_{1},...,z_{N_{f}})\in |d|^{N_{f}}}^{\infty} \prod_{j=1}^{N_{f}} f_{GW}(z_{j}) \prod_{1<j<l\leqslant N_{f}} | z_{j}-z_{l}|^{2},
\end{eqnarray}
where $d$ is a finite domain, a discrete subset of $\mathbb{R}$ with size $|d|$.

The Gross-Witten matrix integral (\ref{matrix continous}) can be expanded for small values of the eigenvalues around $\alpha =0$ and the leading matrix integral becomes a Gaussian matrix integral with a quadratic potential $V_{QP}(\alpha)={\frac{-\alpha^{2}}{2}}$, as follow
\begin{eqnarray}
\label{Gauss MI}
\Z_{QP}=\prod_{j=1}^{N_{f}} \int_{-\pi}^{\pi} \frac{d\alpha_{j}}{2\pi} f_{QP}(\alpha_{j}) \prod_{l<p} | \alpha_{l}-\alpha_{p}|^{2},
\quad
f_{QP}(\alpha)=e^{tV_{QP}}.
\end{eqnarray}
Similar to Eq. \ref{matrix discrete}, the discrete version of the above matrix integral (\ref{Gauss MI}) can be written easily. Later we will discuss the importance and implication of the Gaussian matrix integral in the asymptotic limit of the correlation functions in XX0 model. In fact, we will find that a partition function for the XX0 model with weakly coupled ($\Delta \ll 1$) Hamiltonian is equal to the Gaussian matrix integral.

Using the Heine-Szeg\"{o} identity \unscite{Heine}, which identifies the matrix integrals with Toeplitz and Hankel determinants, we can replace the partition functions of infinite and finite size XX0 models, represented as continuous and discrete matrix integrals Eqs. 4-6, with the continuous and discrete Toeplitz Eq. \ref{Teoplitz}, and Hankel determinants Eq. \ref{Hankel}, as follow
\begin{eqnarray}
\label{Teoplitz Z}
\Z_{GW}=D_{N_{f}}(f_{GW}),\quad  \Z_{GW}^{|d|}=D_{N_{f}}^{|d|}(f_{GW}),
\end{eqnarray}
and
\begin{eqnarray}
\label{Hankle Z}
\Z_{QP}=\mathcal{H}_{N_{f}}(f_{QP}),\quad  \Z_{QP}^{|d|}=H_{N_{f}}^{|d|}(f_{QP}).
\end{eqnarray}

The reviewed results in this section are the key steps toward the explicit calculations of the partition functions and free energy of the (in)finite XX0 spin chain. As we will observe in the following section, the  explicit form of free energy will be used to extract the phase transitions in the model and also to interpret the physical meaning of the different domains in the phase structure. In fact, we will use the recent results in NIBM \unscite{Baik} in a twofold way. First, we use the correspondence between dynamics of magnons in the ferromagnetic vacuum and the motion of nonintersecting walkers in the NIBM. This correspondence pave the way to prove that the partition function of XX0 model can be written as partition function (Toeplitz/Hankel determinants) of a NIBM. Second, we use exact results for the probability of the width of random walkers in the NIBM which is expressed in terms of the Toeplitz and Hankel determinants. The distribution of this width is the Tracy-Widom function which has different trends in its tails. This difference is the origin of phase transitions in the XX0 model.

\section{New results for free energy and phase structure of XX0 Model}
\label{s4}
In this section, our goal is to find explicit expression for the free energy of XX0 model which determines phase structure of the model. Later, by using the exact correspondence between NIBM and XX0 model, we interpret the obtained results.

Let us quickly overview the important methods and new results of this study. Our main strategy to obtain the free energy and phase structure of XX0 model is first to use the matrix integral representation of the XX0 partition function and second to apply the exact results of NIBM, to XX0 model by using the connections between XX0 model and NIBM. In precise words, by using the two main results from \unscite{Bogoliubov},\unscite{Baik}, reviewed in previous chapter and in Appendix C, we obtain our new result indicating that the ratio of partition function of finite and infinite XX0 model, in the asymptotic limit, is given by the Tracy-Widom distribution function, see Eq. \ref{Z/Z F Gw}. A careful asymptotic analysis of this result gives us an explicit new formulas for free energy of finite XX0 model, in different regions of the moduli space, see Eqs. \ref{free enrgy1} and \ref{free enrgy2}. Based on this, we extract the new phase diagram of the XX0 model, which indicates the existence of second- and third-order phase transitions in XX0 model, see Fig. \ref{n_tau}. Finally, we provide a possible interpretation for the new phase structure of the model by assigning a quantum state to each region in the phase diagram, see Fig. \ref{3ph_new_GW}.

In order to carefully translate the results from NIBM to XX0 model we use a dictionary, that relates the parameters of two models. In this study, following \unscite{Bogoliubov}, and as we mentioned before, \textit{N}, \textit{$N_{f}$} and \textit{t} are length of ferromagnetic vacuum, length of the flipped spins (number of magnons) and time, respectively. In the NIBM, \textit{N}, \textit{$N_{f}$} and \textit{t} are the size of system, number of vicious walkers and time, respectively. And in the random matrix theory, \textit{N}, \textit{$N_{f}$} and \textit{t} are the indicator of the discrete structure, the rank of the random matrix and a parameter in the weight function, respectively.

Now, we can explain new fundamental result of this study. We adopt a procedure to determine the asymptotic free energy of the finite XX0 model. In fact, instead of performing a direct asymptotic analysis of discrete and continuous partition functions, i.e. the discrete and continuous Toeplitz and Hankel determinants, we use the obtained results for the ratio of discrete and continuous determinants in terms of probability distributions in the context of NIBM. This mathematical approach has been also used in the context of Chern-Simons theory \unscite{zahabi}. 
In other words, first we use the Toeplitz/Hankel determinant representation of the partition function of XX0 model,  Eq.~\ref{Teoplitz Z} and Eq.~\ref{Hankle Z}, and second we obtain the asymptotic of the discrete and continuous Toeplitz/Hankel determinants from the asymptotic of the probability distributions of the width in NIBM.

In the following, we consider the Gross-Witten case with weight function $f(z)=e^{\frac{t}{2}(z+z^{-1})}$. Combining Eqs.~\ref{con prob1} and \ref{con prob2} in Appendix \ref{MIG}, and using
\begin{equation}
\frac{D_{N_{f}}^{|d_{s}|}(f,|d_{s}|)}{D_{N_{f}}(f)}=\frac{\Z_{GW}^{|d|}}{\Z_{GW}},
\end{equation}
one can obtain
\begin{eqnarray}
\label{integral equation}
\lim_{min(N,N_{f},t)\to\infty}\oint_{|s|=1} \frac{\Z_{GW}^{|d|}}{\Z_{GW}} \frac{ds}{2\pi i s}= F(\frac{N-\mu}{\sigma}),
\end{eqnarray}
where $\mu$ and $\sigma$ are defined by
\begin{equation}
 \label{mu-sigma new}
 \mu  \coloneqq
  \begin{cases}
   N_{f}+t & \quad N_{f} \geq t\\
   2\sqrt{N_{f}t} & \quad N_{f} < t\\
  \end{cases},
   \quad
 \sigma  \coloneqq
  \begin{cases}
   2^{-\frac{1}{3}}t^{\frac{1}{3}} &  \quad N_{f} \geq t\\
   2^{-\frac{2}{3}}t^{\frac{1}{3}}(\sqrt{\frac{N_{f}}{t}}+\sqrt{\frac{t}{N_{f}}})^{\frac{1}{3}} & \quad N_{f} < t\\
  \end{cases}.
\end{equation}
In case $s=1$, Eq. \ref{integral equation}, gives the following new result for XX0 spin chain,
\begin{eqnarray}
\label{Z/Z F Gw1} 
\lim_{min(N,N_{f},t)\to\infty}\frac{\Z_{GW}^{|d|}}{\Z_{GW}}= F(\frac{N-\mu}{\sigma}).
\end{eqnarray}
In the same manner, one can obtain similar result for the quadratic potential. In summary, by combining the results for probability distributions of NIBM, in Appendix \ref{MIG}, separately in  Eqs.~\ref{con prob1} and \ref{con prob2} for $s=1$, and in Eqs.~\ref{Hankel con prob1} and \ref{Hankel con prob2} for $s=0$ and $t=N_f$, through an appropriate re-scaling and limiting process we obtain that the ratio of Toeplitz/Hankel determinants in the double scaling limit is the Tracy-Widom distribution and thus one can obtain the following new and fundamental results in XX0 spin chains, which is a concrete relation between partition functions of finite and infinite XX0 spin chains, 
\begin{eqnarray}
\label{Z/Z F Gw} 
\lim_{min(N,N_{f},t)\to\infty}c_{GW/QP}\frac{\Z_{GW/QP}^{|d|}}{\Z_{GW/QP}}=F(x_{GW/QP}),
\end{eqnarray}
where $F(x)$ is Tracy-Widom distribution function \unscite{Tracy} (see Appendix B), and in the case of Gross-Witten potential, $c_{GW}=1$ and $x_{GW}=\frac{N-\mu(N_f,t)}{\sigma(N_f,t)}$
and for the quadratic potential, $c_{QP}=(N\sqrt{N_f}/\sqrt2\pi)^{-N_f}$ and $x_{QP}=(N-2\sqrt{N_{f}})2^{2/3}N_{f}^{1/6}$. Equation \ref{Z/Z F Gw} explains how the partition function of finite XX0 spin chain is related to the infinite one by the Tracy-Widom distribution function. All the new results for the free energy and phase structure of XX0 spin chain in the following sections, are direct and indirect consequences of this result after some re-scaling.

\subsection{Free energy and phase structure of XX0 spin chain; Gross-Witten case}
In this part, by using our new result, Eq. \ref{Z/Z F Gw} in the Gross-Witten case we will derive the explicit form of the free energy of finite XX0 spin chain in the asymptotic limit and we will extract the phase diagram of the model.

In order to calculate the free energy of the finite XX0 model, we need first the free energy of the infinite spin chain, then by using  Eq.~\ref{Z/Z F Gw} we can obtain the free energy of the finite spin chain.
 The free energy of infinite XX0 model, defined as ${\F}_{GW}=\frac{1}{N_{f}^{2}}\log \Z_{GW}$, can be expressed in terms of the continuous matrix integral with the Gross-Witten potential, Eq. \ref{matrix continous}. In this way, the free energy in the asymptotic limit is obtained first in \unscite{Gross} and then rigorously proved in \unscite{johansson}, as follow
\begin{equation}
 \label{GW PhT}
 \lim_{N_{f},t\to\infty}{\F}_{GW}(\tau) = 
  \begin{cases}
   \frac{\tau^{2}}{4} & \quad 0 < \tau \leq 1  \\
   \\
   \tau-\frac{{3}}{4}-\frac{{\log \tau}}{2} & \quad \tau > 1 \\
  \end{cases},
\end{equation}
 where the ratio $\tau=\frac{t}{N_f}$ is fixed in the asymptotic limit $N_f,t\rightarrow\infty$. It can be observed that there is a discontinuity in third order derivative of free energy ${\F}_{GW}$ at $\tau=1$, which indicates a third-order phase transition in the infinite XX0 spin chain. This phase transition has been noticed and studied in \unscite{perez} and \unscite{bogolyobovRev}, and possible implications of that have been discussed. Furthermore, there are similar phase transitions in NIBM \unscite{forrester} and other relevant combinatorial models such as tilings of Aztec diamonds \unscite{colomo}. The third-order phase transitions are recently reviewed in \unscite{Schehr}. We will provide a new physical interpretation of this phase transition in terms of the diffusion of magnons in the ferromagnetic vacuum in the Sec.~\ref{s5}.

Having an explicit form of the free energy of infinite XX0 model, Eq. \ref{GW PhT}, now we are in position to apply the methods from \unscite{zahabi} and explicitly calculate the free energy of finite XX0 model by using Eq.~\ref{Z/Z F Gw}. This equation can be rewritten for the free energy of finite XX0 model as 

\begin{eqnarray}
\label{f+f=F} 
\lim_{N,N_{f},t\to \infty }\F_{GW}^{|d|}=\lim_{N,N_{f},t\to \infty}\left(\F_{GW}+\frac{1}{N_{f}^{2}}\log  F(x)-\frac{1}{N_{f}^{2}}\log c_{GW}\right),
\end{eqnarray}
where $\F_{GW}^{|d|}=\frac{1}{N_{f}^{2}}\log \Z_{GW}^{|d|}$ and $\textit{x}$, the argument of Tracy-Widom distribution, is  $x=\frac{N-\mu(N_{f},t)}{\sigma(N_{f},t)}$. By fixing the following parameters; rescaled time $\tau=\frac{t}{N_f}$ and inverse magnon density $n^{-1}=\frac{N}{N_f}$, the argument of the Tracy-Widom distribution can be written as $x=j N_{f}^{2/3}$ where 
\begin{equation}
 \label{j}
 j=  
  \begin{cases}
   \frac{n^{-1}-(\tau+1)}{2^{-\frac{1}{3}}\tau^{\frac{1}{3}}}  & \quad \tau \leq 1   \\
   \\
   \frac{n^{-1}-2\tau^{\frac{1}{2}}}{2^{-\frac{2}{3}}\tau^{\frac{1}{3}}(\tau^{\frac{1}{2}}+\tau^{-\frac{1}{2}})^{\frac{1}{3}}}  & \quad \tau > 1  \\
  \end{cases}.
\end{equation}
Also notice that in the asymptotic limit, $N,N_{f},t \rightarrow \infty$, for $N>\mu$, \textit{x} tends to $\infty$ and for $N<\mu$, \textit{x} tends to $-\infty$ whereas the inverse density $n^{-1}$ at the critical point ($N=\mu$), can be written by using Eq.~\ref{mu-sigma new}, as
\begin{equation}
 \label{mu-sigma-mod}
 n^{-1}=\frac{N}{N_{f}}=\frac{\mu}{N_{f}}  =
  \begin{cases}
   1+\tau & \quad \tau \leq 1\\
   2\sqrt{\tau} & \quad \tau > 1\\
  \end{cases}.
\end{equation}
Hence, the asymptotic behavior of the Tracy-Widom distribution, Eq. \ref{ATW} in Appendix B,  
\begin{equation}
 F(x) = 
  \begin{cases}
   1-\mathcal{O}(e^{-x^{3/2}}) & \quad x \rightarrow \infty \\
   \\
   \mathcal{O}(e^{-|x|^{3}}) & \quad x \rightarrow -\infty \\
  \end{cases},
\end{equation}
can be written in terms of the XX0 models parameters as 
\begin{equation}
\label{domain wall}
 F(n^{-1},t) = 
  \begin{cases}
   1-\mathcal{O}\big(e^{-j^{3/2}N_{f}}\big) & \quad n^{-1}>1+\tau \quad   , \quad \tau \leq 1\\
   1-\mathcal{O}\big(e^{-j^{3/2}N_{f}}\big) & \quad n^{-1}>2\sqrt{\tau}\quad  , \quad \tau > 1\\
   \mathcal{O}\big(e^{-|j|^{3}N_{f}^{2}}\big) & \quad n^{-1}<1+\tau \quad  ,\quad \tau \leq 1\\
   \mathcal{O}\big(e^{-|j|^{3}N_{f}^{2}}\big) & \quad n^{-1}<2\sqrt{\tau}\quad  , \quad \tau > 1\\
  \end{cases},
\end{equation}
where $j$ depends on $\tau$ as in Eq. \ref{j}. As we will see, the asymptotic behavior of the Tracy-Widom distribution is the deriving source for the phase structure of the XX0 spin chain and in fact the conditions in Eq. \ref{domain wall} determine the domain walls in the phase diagram of the model.
Finally, Substituting  Eq.~\ref{GW PhT} and precise form of Eq.~\ref{domain wall} (by using Eq. \ref{asym TW}) in Eq.~\ref{f+f=F}, we can write the free energy of the finite XX0 mode in the asymptotic limit as
\begin{equation}
 \label{free enrgy1 log}
 for\quad \tau \leq 1 :\lim_{N,N_{f},t\to\infty}\mathbb{\F}_{GW}^{|d|}=  
  \begin{cases}
   \frac{\tau^{2}}{4}+\lim_{N_{f}\to\infty}\bigg(\frac{1}{N_{f}^{2}}\log (1-\frac{e^{-c_{1}j^{3/2}N_{f}}}{32\pi j^{3/2}N_{f}})\bigg) & \quad  n^{-1} > \tau+1 \\
   \\
   \frac{\tau^{2}}{4}+\lim_{N_{f}\to\infty}\bigg(\frac{1}{N_{f}^{2}}\log (c_{3}\frac{e^{-c_{2}|j|^{3}N_{f}^{2}}}{|j|^{1/8}N_{f}^{1/12}})\bigg) &  \quad n^{-1} < \tau+1\\
  \end{cases},
\end{equation}
where $c_{1} = 4/3$, $c_{2} = 1/12$ and $c_{3}=2^{\frac{1}{42}e^{\xi(-1)}}$ 
and
\begin{equation}
 \label{free enrgy2 log}
 for\quad \tau > 1 :\lim_{N,N_{f},t\to\infty}\mathbb{\F}_{GW}^{|d|}=  
  \begin{cases}
   \tau-\frac{{3}}{4}-\frac{{\log \tau}}{2}+\lim_{N_{f}\to\infty}\bigg(\frac{1}{N_{f}^{2}}\log (1-\frac{e^{-c_{1}j^{3/2}N_{f}}}{32\pi j^{3/2}N_{f}})\bigg) & \quad n^{-1} > 2\sqrt \tau \\
   \\
   \tau-\frac{{3}}{4}-\frac{{\log \tau}}{2}+\lim_{N_{f}\to\infty}\bigg(\frac{1}{N_{f}^{2}}\log (c_{3}\frac{e^{-c_{2}|j|^{3}N_{f}^{2}}}{|j|^{1/8}N_{f}^{1/12}})\bigg) & \quad  n^{-1} < 2\sqrt \tau\\
  \end{cases}.
\end{equation}
After expanding the logarithm in Eqs.~\ref{free enrgy1 log} and \ref{free enrgy2 log} and keeping the finite terms in the asymptotic limit, the explicit form of the free energy of finite XX0 model can be obtained as 
\begin{equation}
 \label{free enrgy1}
 for\quad \tau \leq 1 :\lim_{N,N_{f},t\to\infty}\mathbb{\F}_{GW}^{|d|}=  
  \begin{cases}
   \frac{\tau^{2}}{4}  & \quad n^{-1} > \tau+1 \\
   \\
   \frac{\tau^{2}}{4}-\frac{c_{2}}{2^{-1}\tau}|n^{-1}-(\tau+1)|^{3}  &  \quad n^{-1} < \tau+1\\
  \end{cases},
\end{equation}
and
\begin{equation}
 \label{free enrgy2}
 for\quad \tau > 1 :\lim_{N,N_{f},t\to\infty}\mathbb{\F}_{GW}^{|d|}=  
  \begin{cases}
   \tau-\frac{{3}}{4}-\frac{{\log \tau}}{2} & \quad n^{-1} > 2\sqrt \tau \\
   \\
   \tau-\frac{{3}}{4}-\frac{{\log \tau}}{2} -\frac{c_{2}}{2^{-2}\tau(\tau^{\frac{1}{2}}+\tau^{-\frac{1}{2}})}|n^{-1}-2\sqrt \tau|^{3} & \quad n^{-1} < 2\sqrt \tau\\
  \end{cases}.
\end{equation}

\begin{figure}[t]
\begin{center}
\scalebox{0.5}{\includegraphics[angle=0]{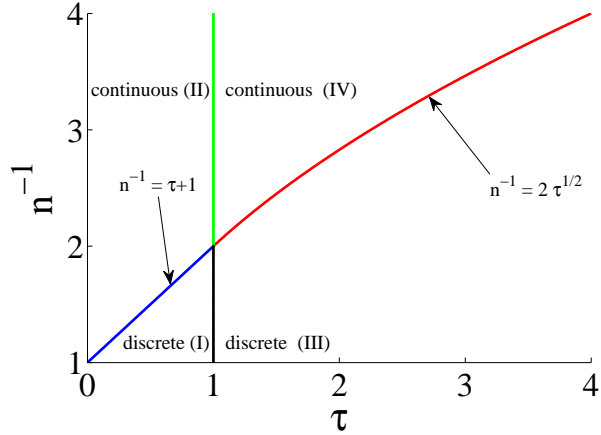}}
\end{center}
\caption{Phase structure of finite XX0 model plotted on inverse density (\textit{n}) versus rescaled time ($\tau$) diagram. Blue (border of region I \& II), green (border of region II \& IV) and 
red (border of region III \& IV) lines are third-order phase transition domain walls while black line (border of region I \& III) is a 
second-order domain wall.
}
\label{n_tau}
\end{figure}

The above explicit results, Eqs. \ref{free enrgy1} and \ref{free enrgy2}, indicate and determine the complete phase structure of the XX0 model. As it can be seen from the four conditions in Eqs. \ref{free enrgy1} and \ref{free enrgy2}, the phase diagram of the model has four domain walls dividing the diagram to the four regions. The order of discontinuity of derivative of free energy at each domain wall determines the order of each phase transition in the phase diagram. The phase diagram of free energy is plotted in Fig.~\ref{n_tau}, in which the second- and third-order phase transitions can be observed.
In fact, the red, green and blue lines are third-order domain walls, and the black line is a second-order domain wall. 
A possible physical interpretation of these phase transitions in the XX0 model will be discussed in Sec.~\ref{s5}. As we will explain, the discrete/continuous label in different regions of phase diagram, Fig. 1, refers to the fact that the XX0 model is described by discrete/continuous matrix model in that region of the moduli space.

\subsection{Free energy and phase structure of XX0 spin chain; quadratic potential case}
As we have seen in chapter two, the Gross-Witten matrix integral representation of the XX0 model in the leading approximation, for small values of $\alpha$, reduces to a matrix integral with the quadratic potential, Eq. \ref{Gauss MI}. In the first step, we will elaborate on the physical meaning of this quadratic/Gaussian matrix integral from the XX0 model point of view. Then, we will follow the same procedure as previous section and we obtain new results for the free energy of this model and its phase structure. 

To understand the physical meaning of the Gaussian matrix model for XX0 model, we investigate on a possible minimum deformation in the parameters the partition function of XX0 model,
that reflects the change of the weight function in the matrix integral from the Gross-Witten potential to the quadratic potential. However, in our approach we only look for a deformation that does not change the structure of the Hamiltonian of the model. The Gaussian matrix integral is the leading term in the $\alpha$-expansion of the Gross-Witten matrix integral and thus it approximately explains the XX0 spin chain in some limit. The Gaussian matrix model gives the leading term of the partition function of XX0 model, which is on the other hand given by $\Delta$-expansion of the spin average of the time evolution operator in Eq. \ref{Delta eq7}. Therefore,  although $\alpha$- and $\Delta$-expansion are obviously different and independent, we can interchangeably use these expansions and approximate the leading term of the XX0 partition function in the $\Delta$-expansion by the Gaussian matrix integral. Moreover, the leading term in the $\Delta$-expansion of Eq. \ref{Delta eq7} can be considered as the partition function of the XX0 model when we assume $\Delta \ll 1$. Therefore, in comparison with the strong coupling case, $\Delta = 1$, in which the XX0 model is described by Gross-Witten matrix integral, a weakly coupled XX0 model and its partition function are governed by the Gaussian matrix integral. In fact, this is the only possible interpretation of the Gaussian matrix integral for the spin chain model with the same XX0 Hamiltonian.

Similar to the previous section, we first obtain the free energy of the weakly coupled infinite XX0 model and then by using a version of Eq.~\ref{Z/Z F Gw} for quadratic potential, we can obtain the free energy of the weakly coupled finite XX0 model as our final result. Before that, we must discuss about resolution of a possible issue.
Using the inherited scaling $\tau=\frac{t}{N_f}$ from the Gross-Witten case, we can replace the parameter $t$ by $\tau N_f$ in the quadratic potential in Eq. \ref{Gauss MI}. Then, we should determine the value of $\tau$. It is easy to see that, if we choose $\tau\searrow 1$, ($\tau$ tends to one in the regime $\tau>1$) as $t,N_f\rightarrow \infty$, then by using Eq. \ref{mu-sigma new} in the regime $t>N_f$, $x_{GW}=\frac{N-\mu}{\sigma}$ up to a numerical factor tends to $x_{QP}=(N-2\sqrt{N_f})2^{2/3}N_f^{1/6}$, provided that $N$ in the Gross-Witten case is replaced by $N\sqrt{N_f}$ in the quadratic potential case. This will be explained and interpreted at the end of this section. Thus, if we fix $\tau=1+\epsilon$ (for infinitesimal parameter $\epsilon$), the scaling behavior of the XX0 model with Gross-Witten potential becomes that of the XX0 model with the quadratic potential. Thus, $\tau=1+\epsilon$ is a right choice to transit from Gross-Witten case to Gaussian case. 
Moreover, for $\tau=1+\epsilon$ the matrix integral \ref{Gauss MI} becomes the Gaussian matrix integral with the weight function $f=\exp{(-N_f\frac{\alpha^2}{2})}$ and this has a convenient form for applying the known mathematical results from \unscite{Baik}. 
     
In the asymptotic limit, the free energy of the infinite system can be easily obtained from ${\F}_{QP}=\frac{1}{N_{f}^{2}}\log \Z_{QP}$ by using the exact expression for the matrix integral with quadratic potential Eq. \ref{Gauss MI}, as $t,N_f\rightarrow\infty$ with fixed $\tau=t/N_f=1+\epsilon$. This exact expression is called Selberg integral, \unscite{Mehta},\unscite{Mehta1}, and it can be read as
\begin{eqnarray}
\label{G Free energy}
\Z_{QP}=(\tau N_f)^{-N_f}\prod_{j=1}^{N_{f}} \int_{-\infty}^{\infty} \frac{dx_{j}}{2\pi} e^{-x_j^2/2} \prod_{l<p} | x_{l}-x_{p}|^{2}=(\tau N_f)^{-N_f}\prod_{j=1}^{N_f}\frac{\Gamma(1+j)}{\Gamma(2)}.
\end{eqnarray}
The free energy of the weakly coupled finite XX0 model in the asymptotic limit is consequently obtained from Eq.~\ref{Z/Z F Gw} as
\begin{eqnarray}
\label{G free energy0} 
\lim_{N,N_{f},t\to \infty }\F_{QP}^{|d|}=\lim_{N,N_{f},t\to \infty}\left(\F_{QP}+\frac{1}{N_{f}^{2}}\log  F(x)-\frac{1}{N_{f}^{2}}\log c_{QP}\right),
\end{eqnarray}
where ${\F}_{QP}^{|d|}=\frac{1}{N_{f}^{2}}\log \Z_{QP}^{|d|}$ and  $x=(N-2\sqrt{N_{f}})2^{2/3}N_{f}^{1/6}$. Let us fix 
$\lambda=\frac{N}{\sqrt{N_{f}}}$ in the asymptotic limit, then the argument of the Tracy-Widom distribution becomes $x= j N_{f}^{2/3}$ 
with $j=2^{2/3}(\lambda-2)$. 
In the asymptotic limit, using similar argument to one in the previous section, the sign of \textit{x} at infinity depends on the value of $\lambda$ (relative value of $N$ and $\sqrt{N_f}$) and the asymptotic behavior of the Tracy-Widom distribution, Eq. \ref{ATW} in Appendix B, can be read as
\begin{equation}
 \label{asym TW inf}
 F(N,N_f) = 
  \begin{cases}
   1-\mathcal{O}\big(e^{-j^{3/2}N_{f}}\big) & \quad N > 2\sqrt{N_{f}} \\
   \mathcal{O}\big(e^{-|j|^3N_{f}^{2}}) & \quad N < 2\sqrt{N_{f}} \\
  \end{cases}.
\end{equation}
Finally, substituting  Eq.~\ref{G Free energy} and precise version of Eq. \ref{asym TW inf} in Eq.~\ref{G free energy0}, we can write the free energy of weakly coupled finite XX0 model in the asymptotic limit, explicitly as
\begin{eqnarray}
 \label{G free enrgy2 log}
 \lim_{N,N_{f}\to\infty}{\F}^{|d|}_{QP}=  
  \begin{cases}
   \lim_{N_{f}\to\infty}\bigg({\F}_{QP}+\frac{1}{N_{f}^{2}}\log (1-\frac{e^{-c_{1}j^{3/2}N_{f}}}{32\pi j^{3/2}N_{f}})-\frac{1}{N_{f}^{2}}\log(\frac{\lambda N_f}{\sqrt2\pi})^{-N_f}\bigg) & \quad N > 2\sqrt{N_{f}} \\
   \\
   \lim_{N_{f}\to\infty}\bigg({\F}_{QP}+\frac{1}{N_{f}^{2}}\log (c_{3}\frac{e^{-c_{2}|j|^{3}N_{f}^{2}}}{|j|^{1/8}N_{f}^{1/12}})-\frac{1}{N_{f}^{2}}\log(\frac{\lambda N_f}{\sqrt2\pi})^{-N_f}\bigg) & \quad N < 2\sqrt{N_{f}} \\
  \end{cases}.\nonumber\\
\end{eqnarray}
Keeping the finite leading terms in the large $N_{f}$ limit, we obtain
\begin{eqnarray}
 \label{G free enrgy2}
 \lim_{N,N_{f}\to\infty}{\F}^{|d|}_{QP}=  
  \begin{cases}
   \mathcal{A}_{QP}  & \quad \lambda > 2 \\
   \\
   \mathcal{A}_{QP}-\frac{1}{3}|\lambda-2|^{3}   & \quad \lambda < 2 \\
  \end{cases},
\end{eqnarray}
where $\mathcal{A}_{QP}=\lim_{N_f\rightarrow\infty}\frac{1}{N_f^2}\sum_{j=1}^{N_f}\left(\log\Gamma(1+j)-\log\Gamma(2)\right)$.

\begin{figure}[t]
\begin{center}
\scalebox{0.5}{\includegraphics[angle=0]{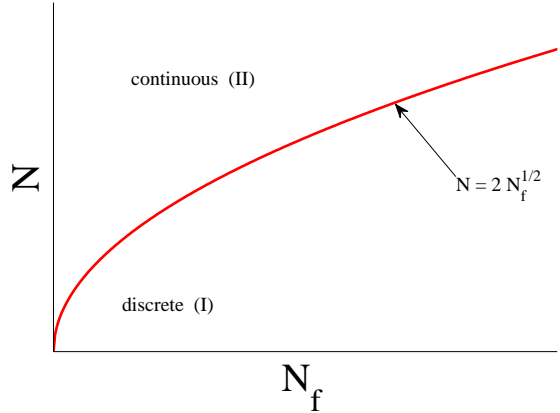}}
\end{center}
\caption{Phase structure of XX0 model in the weak coupling regime plotted on number of magnons $N_{f}$ versus size indicator $N$ diagram. Two phases of the system are divided by a second-order domain wall.  
}
\label{N_Nf}
\end{figure}

Using the free energy of weakly coupled finite XX0 model, Eq. \ref{G free enrgy2}, the phase structure of the model can be extracted as in Fig.~\ref{N_Nf}. It can be easily observed that the red line in Fig.~\ref{N_Nf} is a third-order domain wall, separating two regions in the moduli space of parameters. Using other plausible methods, similar phase transition for the Gaussian matrix model has been observed in the Douglas-Kazakov model, i.e. two-dimensional Yang-Mills theory \unscite{Douglas}. In context of NIBM with periodic, absorbing and reflecting boundary conditions, similar methods of matrix models with Gaussian measure have been applied and by using the Tracy-Widom distribution, similar phase structure is obtained \unscite{Majumdar}, \unscite{Forrester}.     

As final remark in this section, we comment on replacement of $N$ (in Gross-Witten model) by $N\sqrt{N_f}$ (in Gaussian model). First, we should mention that, it is natural to expect different qualitative behaviors of the XX0 model in weak and strong coupling regimes, since they are described by different matrix models with different potentials. Thus, the scaling behavior, $N\sim \sqrt{N_f}$, of the weakly coupled XX0 model (remember that we fix $\lambda=N/\sqrt{N_f}$) is expected to be qualitatively different from that of the strongly coupled model, $N\sim N_f$. However, this different scaling, $N\sim \sqrt{N_f}$, in the weakly coupled model indicates that as far as $N_f$ is interpreted as the number of magnons (which has dimension of size) in the proposed dictionary, $N$ can't be interpreted as the size of the spin chain, anymore. In fact, the appropriate dictionary for the weakly coupled system, can be obtained from the Gaussian matrix integral and its discretization on the domain, $d_s\coloneqq \{ \frac{\sqrt{2}\pi}{N\sqrt{N_{f}}}m|m \in \mathbb{Z} \}$, see Appendix C. From the definition of the domain $d_s$, it can be seen that the size or the largest scale of the system is $N\sqrt{N_f}$. This different scaling in this model can be traced back to the fact that weight function in this case, $f_{QP}=\exp(-N_f\frac{x^2}{2})$, is scaled by $N_f$. Thus, we prescribe that the size of the weakly coupled finite XX0 spin chain is proportional to $N\sqrt{N_f}$. This interpretation is also consistent with conditions in Eqs. \ref{G free enrgy2 log} and \ref{G free enrgy2}, since we interpret the size of the chain to be $N\sqrt{N_f}$, then the number of magnons, $N_f$, can't be larger than the size of the chain, which is naturally expected.

Before we continue with the interpretation of the results, lets us make some comments about the new results of this study. The phase structures of XX0 model and its weak coupling limit, Fig. 1 and 2 are new results of this study in the context of spin chains. The new mathematical method and the careful detailed asymptotic analysis, to obtain the phase structure of matrix models is developed by one of the authors in the context of Chern-Simons theory, \unscite{zahabi}, and our purpose in this study is first to extend this to other matrix models with different potentials such as quadratic potential and more importantly to apply this in the context of the XX0 spin chain and extract the phase structure of this model. Furthermore, as we will see in the next section, we obtain and present a genuine interpretation of the obtained mathematical results for the phase structure of XX0 model. Thus, in this study we obtain not only a new phase structure for the XX0 model, but we also explore the physical meaning and interpretation of the phase structure.

To compare with the similar results obtained with different plausible methods in similar contexts, we mention the study in \unscite{perez}, in which the phase transition in the XX0 model is partially conjectured only for the third-order phase transition in infinite size model, via its similarities to continuous Gross-Witten model. However, the complete phase structure of the finite size model, including third- and second-order phase transitions and the crucial role of the TW distribution, and also the interpretations of the phase transitions in finite/infinite models are discussed in this study. In a related context, third-order phase transitions in non-intersecting Brownian bridge, Gaussian discrete matrix (Douglas-Kazakov) models, are studied using the TW distribution \unscite{majumdar} and \unscite{Forrester}. Beside the similarities between our analysis of the phase structure in weakly coupled XX0 model and the one for non-intersecting Brownian bridge in \unscite{majumdar} and \unscite{Forrester}, the analysis of the phase structure of the NIBM with the Gross-Witten potential (the original XX0 model) using the TW distribution leads to the new result of this study.

\section{Interpretation of XX0 phase structure}
\label{s5}

In previous chapter, we obtained phase transitions in the asymptotic limit of the finite XX0 spin chain in the strong and weak coupling regimes. In this chapter, we provide a pictorial schematic interpretations for the phase structures, (Fig. 1 and Fig. 2) of the XX0 spin chain. This interpretation is based on the dynamics of XX0 model given by the NIBM, and it consists of snapshots of the quantum states (spin configurations) of the spin chain which are determined by the time evolution of the initial state in the asymptotic limit. The snapshots depend on the parameters of the system and they demonstrate different qualitative features of the model in different regions of the moduli space.

Despite the fact that spin chains are one-dimensional statistical models, the quantum and topological effects make the phase structure of XYZ Heisenbeg model interesting enough for intense studies during this and last century.  There are various quantum and topological phase transitions in spin chains such as the XXZ and XY Heisenberg models, caused by effects of anisotropic couplings and external magnetic field, see for example \unscite{Schollwock},\unscite{Q PhT},\unscite{T PhT}. In contrast, the XX0 model, which is an isotropic Heisenberg model, in zero magnetic field have not attended in any of the previously studied quantum or topological phase transitions. However, in this study, we introduce and investigate new phase transitions in the XX0 model which are not belong to the realm of usual quantum or topological phase transitions. The existence of such new phase transitions have been partially discussed in \unscite{perez} and \unscite{bogolyobovRev}, but in this work we obtain a complete phase structure of the model which not only includes previously introduced phase transition but also contains other new phase transitions in this model. In fact, by performing explicit calculations for the free energies in all regions of the phase diagram we determined all the domain walls and the orders of the phase transitions. Moreover, in this section we will provide a possible new interpretation for the obtained results of the phase structure of the model.

Beside the fact that the new phase transitions in XX0 model are determined by the explicit calculations of the free energy,
the existence of equivalent phase transitions that occur in the NIBM \unscite{Baik}, \unscite{Forrester} confirms our claim and furthermore provide us with an approach toward the interpretation of the results. In fact, by using the correspondence between NIBM and XX0 model, as shown in Eqs.~\ref{action} and \ref{action2} in Appendix A, the phase transitions in NIBM which belong to the classical probabilistic and stochastic behavior of the system, help us to understand the physical meaning of the phase transitions in the XX0 model.

In this study, as we explained in the previous chapter, we determined the phase diagrams of the XX0 model in the strong and weak coupling regimes. Before starting the interpretation, we should mention that in the following interpretations since the results are not dependent of boundary conditions we ignore and relax the periodic boundary conditions and we talk about the finite spin chain on a line segment with boundaries. In other words, the type of boundary conditions on the boundaries of the chain plays no role in this analysis.
Our first comment and interpretation is about the asymptotic limit of the finite XX0 model. In fact, as it can be seen from  Eqs. \ref{free enrgy1}, \ref{free enrgy2} and \ref{G free enrgy2}, the asymptotic limit of the free energy of finite XX0 model in some regions of the moduli space (upper conditions in those equations) is given by that of the infinite model and thus we can claim that in these regions of the moduli space the effect of size is washed out in the asymptotic limit and this limit is the continuum limit and hence the model becomes real infinite (continuous) model with no size effect. On the other hand, in other regions of the moduli space (lower conditions in Eqs.~\ref{free enrgy1},~\ref{free enrgy2} and \ref{G free enrgy2}), although the system is in the asymptotic limit and the size of the system is moderately large, but since  the asymptotic limit of the free energy of finite model, in addition to the free energy of the infinite model, has an extra term, the system still shows the effects of size. Thus, we call the model, finite (discrete) model with finite size effects, in the asymptotic limit. This extra term in free energy can be interpreted as the energy of the finite size effects.

In the strong coupling XX0 model, as we have seen, we have two different kinds of phase transition. The first one is the phase transition between region II and IV in Fig. 1, that happens in the infinite XX0 model and it is described by the continuous matrix model and Eq.~\ref{GW PhT}, and basically has no size parameter. The other phase transitions between region (I \& II), (III \& IV) and (I \& III) happen in the asymptotic limit of the finite XX0 model which is governed by the discrete matrix model and Eqs.~\ref{free enrgy1} and \ref{free enrgy2}. In these phase transitions, as we mentioned before, although the size and other parameters of the system are infinitely large, but there are regions (I and IV) in the moduli space that in these regions (phases) the system still shows the effects of size  whereas in the other regions (II and III), the system behaves like a real infinite XX0 model with no size effect and described by a continuous matrix model. That is the origin of the discrete and continuous labels in the phase diagrams, Fig.~\ref{n_tau} and Fig.~\ref{N_Nf}. In the weak coupling regime, since the free energy is given by Eq.~\ref{G free enrgy2}, similar phase transition between discrete and continuous phases of the model happens, but we don't have a phase transition analogous to the one between regions (II and IV) and/or (I and III) in Fig.~\ref{n_tau}. 

Before explaining each of the above phase transitions, we should mention the following important points. In fact, all the interpretations in this section should be understood in the light of the following points:
\begin{itemize}
\item All the reasoning in the interpretation of the results are based on a simple fact, which is the correspondence between: i) $N_f$ magnons and $N_f$ nonintersecting random walkers, and ii) diffusion dynamics of the magnons in the ferromagnetic vacuum of spin chain and the nonintersecting Brownian motion of the random walkers in a fixed domain.

  \item  Based on dynamics of the NIBM and its translation in terms of the diffusion of magnons into the ferromagnetic vacuum, we will associate a quantum state, i.e. a configuration of spins in each region of the phase diagram which demonstrates an equilibrium state obtained from the time evolution of the initial quantum state, used in Eq. \ref{Delta eq7}, in the asymptotic limit.
  In fact, these configuration are different distributions for positions of the magnons in the ferromagnetic vacuum, in the different regions of the phase diagram. The quantum state of each phase reflects the qualitative feature of that phase. Furthermore, the quantum state is in equilibrium, regardless of the local movement of magnons in the vacuum which is a dynamical process.
  
  \item In this study, all the phase transitions in XX0 model take place in the asymptotic limit, $N,N_f,t\rightarrow \infty$, and they are resulted from the competition of these parameters or the dimensionless rescaled parameters $\tau$ and $n^{-1}$. 
\end{itemize}

\subsection{Strongly coupled infinite XX0 spin chain}
First, we interpret the Gross-Witten phase transition and its implications for XX0 model. The free energy of the infinite XX0 spin chain, Eq.~\ref{GW PhT}, determines a third-order phase transition, separating region II and IV in $\tau-n^{-1}$
phase diagram at $\tau =1$ as shown in Fig.~\ref{n_tau}.
In order to explain this phase transition, 
we focus on the dynamical process of the diffusion of magnons in the ferromagnetic vacuum. 
As mentioned in Sec.~\ref{s2} and explained in Appendix A, the diffusion of magnons into the ferromagnetic vacuum
is entirely equivalent to the dynamical process of random vicious walkers in NIBM.

\begin{figure}
\begin{center}
\scalebox{0.5}{\includegraphics[angle=0]{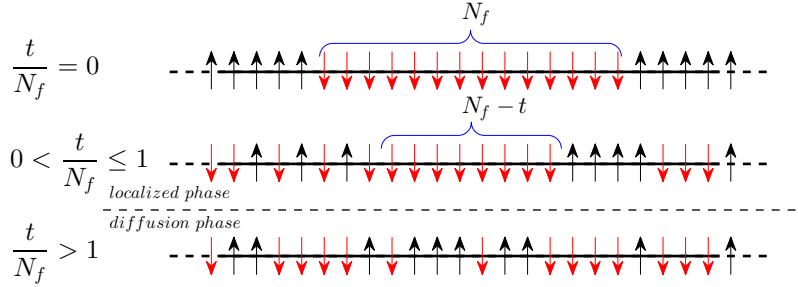}}
\end{center}
\caption{Symmetric version of quantum spin configurations of infinite XX0 model in the region II \& IV of Fig.~\ref{n_tau}. The region II has localized block of magnons and it is called localized phase whereas the region IV is called diffusion phase with all the magnons diffused into the vacuum and it has no localized block of magnons.
}
\label{3ph}
\end{figure}

\begin{figure}
\begin{center}
\scalebox{0.5}{\includegraphics[angle=0]{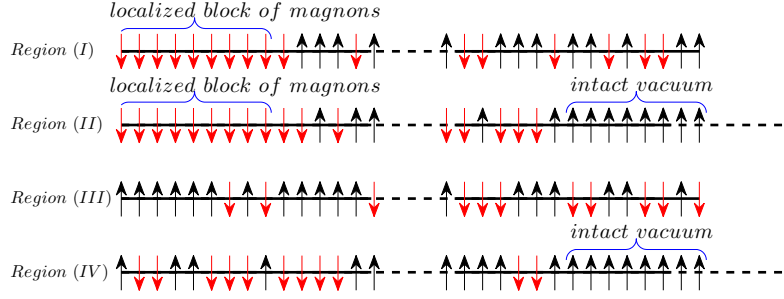}}
\end{center}
\caption{Quantum spin configurations of finite XX0 model in asymptotic limit corresponding to each region of the phase diagram, Fig.~\ref{n_tau}.
}
\label{3ph_new_GW}
\end{figure}

Let us start with a collection of $N_f$ neighbouring magnons, located in an arbitrary interval in an infinitely large ferromagnetic vacuum, as shown in the Fig.~\ref{3ph}. Notice that, for more clarity we considered a symmetric quantum state which is different from the initial quantum state in Eq. \ref{Delta eq7}.
In this section, we study the time evolution of the symmetric quantum state. 
The diffusion of magnons into the vacuum starts at $t=0$, and take place one spin after the other in each step of time. 
At the beginning of time, there are $N_{f}$ neighboring magnons, called \textit{localized (freezed) block of magnons} with size $N_f$. 
As the system evolves with time, those magnons which are located at the edge of the localized block start to detach from the block
and diffuse to the ferromagnetic vacuum, similar to random vicious walkers in NIBM. As a result, size of the magnon block shrinks with time. 
Since at any step of time the foremost magnons at the two ends of block have a chance to detach from the block, thus at time $t$, length of the magnon block with no chance to diffuse is $N_f-t$.  The process of detaching of magnons from the block would end when $N_f=t$, once all of $N_f$ magnons have the chance to diffuse and the magnon block would disappear.

As we will see now, the above explanation of the dynamics becomes more accurate in the limit of the large parameters, $N_f,t\rightarrow \infty$ and we can explain the phases with more certainty. Now consider the asymptotic limit, according to Eq.~\ref{GW PhT}, depending on the ratio $\frac{t}{N_f}$, be smaller or bigger than one, the system asymptotes to two different quantum states. As a matter of fact, the domain wall, $\tau=1$, separates the phase diagram into two phases, $\tau \leq 1$  and $\tau>1$, with different free energy.
In $\tau \leq 1$ phase, there is a localized block of magnons with non-zero size and the remaining magnons in this block has no chance to diffuse in the vacuum and thus the XX0 spin chain has a long range order in this phase. In fact, this localized block of magnons phase has been already noticed in the context of NIBM, in which a part of vicious walkers path is frozen in $\tau\leq 1$ phase , see Figure 2 in \unscite{Baik}.
However, in $\tau > 1$ phase, all the magnons diffuse into the vacuum and thus there is no localized block of magnons. Equivalently, the XX0 spin chain in this phase has no long range order. 

\subsection{Strongly coupled finite XX0 spin chain}
Having examined and interpreted the Gross-Witten phase transition in infinite XX0 spin chain, now we focus on studying and interpreting the complete phase structure of finite XX0 model, Fig.~\ref{n_tau}.  

The finite XX0 model in the asymptotic limit contains the Gross-Witten phase transition as well as other phase transitions. In the same spirit as previous part, the interpretation of the new phase transitions is based on the diffusion of magnons in the ferromagnetic vacuum and the propagation of the corresponding vicious random walkers. Here, the difference with the previous part is that the XX0 model has the size parameter $N$ which plays a crucial role along the effects of the other parameters, in the asymptotic limit. Thus, as we also mentioned in the beginning of this chapter, the main difference between the asymptotic limit of the finite XX0 model and the infinite model is that in the asymptotic limit, the finite system bifurcates into two chambers of the moduli space with respect to the behavior of the parameter $N$. One is with real infinitely large size XX0 model with no size effect and the other one is with moderately large size XX0 model but still with non-negligible finite size effect.

From a mathematical point of view, the diffusion of the magnons into the vacuum is equivalent to the propagation of vicious walkers in a fixed domain and thus we can study the size effects in the process of magnon diffusion by investigating the probability distribution of the width of the NIBM in a finite domain, see Appendix \ref{MIG}. The size effect in the asymptotic limit of the spin chain is reflected in the dynamics of the magnons and, as we will explain more in this section, we can interpret the effect of size by the event that at least a diffusing magnon (the first magnon in the initial block of magnons next to the vacuum) in the vacuum of the strongly coupled XX0 model reaches to the boundary of the spin chain, in the asymptotic limit. In NIBM, this is equivalent to the event that the width $W_{N_f}$ of the NIBM becomes equal or greater than the size of the domain, $N$. Let us denote the first magnon in the block of magnons by $m$ and the boundary of the spin chain by $B$, and define $m\rightsquigarrow B$ as an event that the first magnon reaches to the boundary of the spin chain. Then by using the fact that the dynamics of the magnons in the spin chain is equivalent to that of the vicious walkers in NIBM, we can write
\begin{equation}
\mathbb{P}(m\rightsquigarrow B)\equiv\mathbb{P}(W_{N_f}\geq N)=1-\mathbb{P}(W_{N_f}<N),
\end{equation}
and thus by using Eqs. \ref{con prob1} and \ref{con prob2} in \ref{MIG}, we obtain
\begin{equation}
\lim_{N,N_f,t\rightarrow\infty}\mathbb{P}(m\rightsquigarrow B)=1-F\left(\frac{N-\mu(N_f,t)}{\sigma(N_f,t)}\right),
\end{equation}
where $N$ is the size of the spin chain. 
Then, by using the asymptotic of the Tracy-Widom distribution, Eq. \ref{domain wall}, we obtain the leading order probability that a magnon reaches to the boundary, 
\begin{equation}
\lim_{N,N_f,t\rightarrow\infty}\mathbb{P}(m\rightsquigarrow B)=
\begin{cases}
   0 & \quad n^{-1}>1+\tau \quad   , \quad \tau \leq 1\\
   0 & \quad n^{-1}>2\sqrt{\tau}\quad  , \quad \tau > 1\\
   1 & \quad n^{-1}<1+\tau \quad  ,\quad \tau \leq 1\\
   1 & \quad n^{-1}<2\sqrt{\tau}\quad  , \quad \tau > 1\\
  \end{cases}.
\end{equation}
This result determines that, in each region of the moduli space of the parameters, whether a diffusing magnon into the vacuum of strongly coupled XX0 model reaches to the boundary of the spin chain (thus we have a finite size effect) or not. 
However, for XX0 model with periodic boundary conditions and no apparent boundaries, we can rephrase the interpretation by saying that, in one phase diffusing magnons can move globally and get to the distances of size of the chain, i.e. of the order $\mathcal{O}(N)$, and unlike the other phase in which their movements are restricted to the local sizes of order $\mathcal{O}(1)$.

One of our goal in this section is to find an interpretation based on the dynamics of the magnons, for the extra term in the asymptotic limit of the free energy of the finite XX0 model in comparison to that of the infinite XX0 model (compare second lines to first lines in Eq.~\ref{free enrgy1} and Eq.~\ref{free enrgy2}). As we briefly mentioned, to interpret the (non)existence of the finite size effects in terms of the dynamics of magnons, it would be natural to think that in the (in)finite XX0 model the diffusing magnons do (not) reach to the borders of the chain. In other words, the conditions of the model in the asymptotic limit and the competition between scaled time and scaled size, which is given by the relations between the parameters of the model (in the first and second lines of the Eq.~\ref{free enrgy1} and Eq.~\ref{free enrgy2}), determine that whether diffusing magnons can reach to the boundary of the chain or not and thus we have or have not the size effects in the model. Therefore, we can summarize that the extra terms in the free energy  emerge because of the finite size effects and in fact they can be interpreted as the interaction energy of the diffusing magnons with the boundary of the chain.

As a result, in the phase that the asymptotic limit of the finite XX0 model is given by the infinite XX0 model, no matter how far the diffusing magnons can proceed into the vacuum, still there will be an infinitely large   ferromagnetic vacuum which is not occupied by the diffusing magnons. Lets name this undisturbed vacuum as \textit{intact vacuum}. Therefore, the absence or present of finite size effect determines the existence or absence of intact vacuum in quantum state of XX0 model, respectively.
In addition, the phase diagram of the finite XX0 model is also separated by $\tau=1$ domain wall, into two phases which is characterized by the absence or presence of the localized block of magnons, as discussed in section (4.1). Therefore, the different regions in the phase diagram of the finite XX0 model in the asymptotic limit are contrasted from each other by inclusion and exclusion of intact vacuum and localized block of magnons, separately.

In the rest of this section, based on the above interpretations, we will provide a physical description and schematic illustration,  Fig. 4, for each region of the complete phase diagram of XX0 model. To summarize, the phase structure of the finite XX0 model, determined by free energies (Eqs. \ref{free enrgy1} and \ref{free enrgy2}), has four domain walls dividing 
$\tau-n^{-1}$
phase diagram into four regions, shown in the Fig.~\ref{n_tau}, and each region has the following interpretation:

\begin{itemize}
\item In region (I) with $n^{-1}<\tau +1$ and $\tau<1$, the quantum state contains a localized block of magnons and some but not all of the magnons diffuse to the vacuum and sweep the whole ferromagenetic vacuum and at least one of them (the last one in the row) reach to the boundary of the chain and thus no intact vacuum remains. Thus, it is a moderately large but finite size spin chain.

\item In region (II) with $n^{-1}>\tau +1$ and $\tau<1$, similar to region (I) there is a localized block of magnons in the quantum state, however, the diffusing magnons can not probe all the ferromagnetic vacuum and thus there will be an intact vacuum in the spin chain. Thus, it is an infinite spin chain.
 
\item In region (III) with $n^{-1}<2\sqrt{\tau}$ and  $\tau>1$, all the magnons from the initial localized block diffuses to the vacuum and there is no localized block of magnons, furthermore, at least a magnon probes the whole vacuum to its boundary and this causes the absence of the intact vacuum in the quantum state. Thus, it is a finite spin chain with a moderately large size.

\item In region (IV) with $n^{-1}>2\sqrt{\tau}$ and $\tau>1$, the quantum state contains an intact vacuum but no localized block of magnons. Thus, it is an infinitely large spin chain.
 \end{itemize}

\subsection{Weakly coupled XX0 spin chain} 
As we noticed in section (3.2), weakly coupled XX0 model has a different scaling compare to strongly coupled model. This different scaling implies a different dictionary and interpretation for the weakly coupled model. However, as we will see, the weakly coupled XX0 model is governed by roughly the same dynamics as the strongly coupled model, obtained from the NIBM. Possible differences are originated from the different potentials in the corresponding matrix models and different values of the coupling in the Hamiltonian. We obtained the weakly coupled system and its Gaussian matrix model from $z\rightarrow 1$ and $t\rightarrow \infty$ limits of the strongly coupled system with the Gross-Witten matrix model, thus we expect the same dynamical behavior for the diffusing magnons and nonintersecting vicious walkers in these limits.
Moreover, the Gaussian matrix model appears in a version of NIBM, called nonintersecting Brownian bridge (see Appendix C.2) and therefore dynamics of weakly coupled XX0 model is obtained from NIBM.

We start our discussion with an obvious required changes to the proposed dictionary in section 3. In contrast to Gross-Witten case that $N$ is the size of the spin chain, in the quadratic potential case, the size of the weakly coupled spin chain is given by $N\sqrt{N_f}$. Equivalently, the width $W_N$ of the NIBM in the Gross-Witten case should be replaced by $W_N\sqrt{N_f}$, as we transfer to the quadratic potential in the limit of Gross-Witten case, see section (3.2) and appendix C. On the other hand, the interpretations of $N_f$ and $t$ as the number of magnons and time, remain the same as in strongly coupled system. However, in the matrix model description of the weakly coupled XX0 model, $t$ and $N_f$ are at infinite limit, and dynamical parameter of the system, time $t$, is hidden since it is identified with $N_f$, as we fixed $\tau=1+\epsilon$, see the discussion in section (3.2).

In the case of weakly coupled infinite XX0 spin chain with partition function \ref{G Free energy}, the location of the system in the moduli space of the models is on the domain wall $\tau=1+\epsilon$ of the strongly coupled XX0 model and thus there is no Gross-Witten type phase transition for the weakly coupled infinite XX0 spin chain. This can also be seen from the fact that in the nonintersecting Brownian bridge, see Appendix C.2, there is no freezed region for the path of vicious walkers and thus all the magnons are moving around in the ferromagnetic vacuum and we do not have localized block of magnons. Thus, the weakly coupled XX0 model in the asymptotic limit is always in the diffusion phase of the strongly coupled model in Fig. 3.
However, as Eq.~\ref{G free enrgy2} indicates, the weakly coupled finite XX0 model in the asymptotic limit has a third order phase transition. Using the interpretations discussed in previous part, it is easy to guess that the source of this phase transition is the finite size effects of the model.

Similar to the previous section, by using Eqs. C.7 and C.8 in Appendix \ref{MIG}, we can obtain the probability that a diffusing magnon in the vacuum of the weakly coupled XX0 model reaches to the boundary of the system in the asymptotic limit as
\begin{equation}
\lim_{N,N_f\rightarrow\infty}\mathbb{P}(m\rightsquigarrow B)=1-F\left((N-2\sqrt{N_{f}})2^{2/3}N_{f}^{1/6}\right).
\end{equation}
Consequently, the asymptotic of the Tracy-Widom distribution, Eq. \ref{asym TW inf}, determines the leading probability that a magnon reaches to the boundary of the system as follow
\begin{equation}
\lim_{N,N_f\rightarrow\infty}\mathbb{P}(m\rightsquigarrow B)=
  \begin{cases}
   0 & \quad N > 2\sqrt{N_{f}} \\
   1 & \quad N < 2\sqrt{N_{f}} \\
  \end{cases}.
\end{equation}
This result determines, in each region of the moduli space of the parameters, that whether a diffusing magnon into the vacuum of weakly coupled XX0 model reaches to the boundary of the spin chain (thus we have finite size effects) or not.

Roughly the same qualitative arguments about the diffusion dynamics of the magnons can be used to interpret the phase transition in weakly coupled XX0 model, Fig. 2. In fact, the competition between the size of spin chain and the size of initial block of magnons leads to the phase transition between a phase that a diffusing magnon reaches to the boundary and we have size effects and another phase that a diffusing magnon does not reach to the boundary and there is no size effect.
In this transition,
$N-N_{f}$ 
diagram is separated by $N=2\sqrt{N_{f}}$ domain wall into two regions (see Fig.~\ref{N_Nf}).
In general, the physical interpretation of regions I and II in Fig.~\ref{N_Nf} is similar to  
regions III and IV in Fig.~\ref{n_tau}, respectively.
In fact, as $\tau=1+\epsilon$, there is no localized block of magnons in any of the two phases and in region (I) number of magnons is bigger or equal than half-szie of the spin chain and at least a magnon reaches to the boundary of spin chain, thus the quantum state contains no intact vacuum and the system is moderately large spin chain governed by discrete matrix model, whereas in region (II) number of magnons is less than half-szie of the spin chain and none of the magnons can reach the boundary and thus the quantum state contains an intact vacuum, and the system is infinite spin chain governed by continuous matrix model, see Fig. 5.  

\begin{figure}
\begin{center}
\scalebox{0.5}{\includegraphics[angle=0]{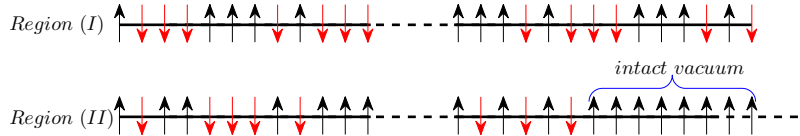}}
\end{center}
\caption{ Quantum spin configurations of weakly coupled XX0 model in different phases of the phase diagram Fig.~\ref{N_Nf}.
}
\label{3ph_new_DK}
\end{figure}

As final remark of this section, we should make it clear that in the proposed equilibrium quantum states, Fig. 4 and 5, which is obtained from the time evolution of the initial quantum states in the asymptotic limit for each phase of the XX0 model, we only focus on the definite features of the phase, such as localized block of magnons and intact vacuum but there are options for the spin configurations in the remaining part of the quantum state. These definite features are permanent characteristics of any quantum states in that phase, but the spin configurations of the remaining parts of the quantum state can be arbitrary chosen. For example, the final quantum state in Eq. \ref{Delta eq7} are chosen to be exactly equal to the initial quantum states. That means all the diffusing magnons should return to their original position and we calculate the probability of this event. In this case, in fact we are calculating the reunion probability for the vicious walkers in NIBM, \unscite{Majumdar}.  From another point of view, we have chosen the initial and final quantum states such that the explicit evaluation of the correlation function, Eq. \ref{Delta eq7}, becomes mathematically feasible, because there is no Schur functions in the matrix integrals, and the explicit results reveal the phase transitions in the XX0 model. However, any other correlation functions in XX0 model, at least with the same initial quantum state as in Eq. \ref{Delta eq7} but with an arbitrary final quantum state, in principle would determine the same phase structure, because the above definite features of the quantum states are still present. For correlation functions with arbitrary initial and final quantum state, see Appendix D, Eq. \ref{C}.

\section{Discussions and future studies}
\label{s6}
In this paper we studied the phase structure of the XX0 model via the explicit formulas for the free energy of the system in different regions of the moduli space of the parameters. The analytic results for the free energy, which led us to the phase structure of the model, were obtained by using the methods from the NIBM and its asymptotic analysis. The Tracy-Widom distribution has played a crucial in this analysis and in fact, the Tracy-Widom distribution was the key tool to obtain the free energy of the system and to extract the phase transitions for the XX0 model.

As we discussed in this article, some exact results about the XX0 spin chain are obtained. These results indicate a definite new phase structure of the model in the the limit of large parameters. It might be imagined that the limit of large parameters in the spin chain is kind of unphysical from the experimental point of view. However, as we have discussed in the interpretation part, very natural explanations for the phase structure in terms of the spin configurations exist. Moreover, some arguments about the possible experimental observations and applications of the similar phase transitions are provided recently in \unscite{perez}.

Having reviewed the obtained results and proposed interpretations, we continue with some possible directions for future studies in this topic.
It might be thought that the appearance of the Tracy-Widom distribution is because of the special characteristics of the model, but in fact, the very existence and appearance of the Tracy-Widom distribution in the asymptotic limit of the NIBM and XX0 model is a general fact, which does not depend on the physical details of the potential and/or other specific physical properties of the model. In fact, the emergence of this distribution in the XX0 model and its generalization is implied by the universality of this distribution in the random matrix theory and NIBM \unscite{baik-suidan}. The Tracy-Widom distribution and hence the phase structure of the XX0 model have a universal character from the mathematical point of view. From physical point of view, one can argue that behavior of the XX0 model near its domain walls (critical points) of the phase transitions is independent of the details of the microscopic interactions and thus, a universal structure for the phase transition is expected. For a review on similar ideas and results, see \unscite{Schehr}. These ideas and facts are important clues for our future studies about universal features of the phase structure of generalized XX0 model with long-range interactions. In fact, concrete mathematical results for the random matrix theory with the generalized form of the Gross-Witten potential \unscite{Baik} can be used to study the generalized XX0 model with infinitely long range interactions. These results indicate the similar phase structure for generalized XX0 model.

In our interpretation, using some mathematical techniques, the heuristic arguments about time evolution of the model and its asymptotic dynamics are obtained. However, the asymptotic dynamics e.g. the diffusion of the magnons into the ferromagnetic vacuum in the asymptotic limit deserves more rigorous studies from the mathematical point of view. The mathematical understanding of this stochastic process seems a feasible problem, starting from the known dynamics for the local motion of the magnons and random walk of the vicious walkers. 

In this study and within our approximation, the dynamics of the system is only governed by the Hamiltonian and we have neglected the effects of quantum fluctuations. However, the quantum fluctuations can play a crucial role in affecting the phase structure of the model, especially in the low temperature limit. The effects of quantum fluctuations become even more important, in the weak coupling regime, in which the aligning force between neighbouring spins is very weak and the intrinsic spin fluctuations become dominant. The quantum fluctuations might affect the purity and stability of the thermal phases discussed in this work, i.e. the stability of localized block of magnons phase and intact ferromagnetic vacuum phase. By stability we mean that any segment of localized block of magnons or intact vacuum that are not adjacent to the spins with opposite direction has stable status and the spins in these segments do not tend to change the direction because of the coupling. Moreover, the competition between the strong/weak couplings and the quantum fluctuations might eventually lead to completely new phase transitions in the system. This requires a separate and comprehensive study about the role of quantum fluctuations in shaping of a complete phase structure of the model.

In this study, we fixed the periodic boundary conditions for the spin chain, however, other boundary conditions such as antiperiodic, reflecting or aborbing are possible and appropriate. The new boundary conditions might bring new possibilities to the picture and deserve further studies. In this sense, some aspects of the problem have been studied in the context of NIBM, \unscite{Majumdar}. 

Another direction for further studies in the phase structure of XX0 model can be explained as follow. The XX0 model and the NIBM are closely related to the combinatorial models such as plane partitions and nonintersecting lattice paths. The enumerative combinatorics of these models provides us with exact techniques for counting the degeneracies in these models which in turn can be used to obtain the entropy of the XX0 model. The entropy consideration opens up the possibility for the investigation of the phase transitions in XX0 model from another perspective and plausible approach.

A possible interesting question about the phase structure of XX0 model is to understand the nature of the second-order phase transition between region I and III in Fig.1 and Fig.4. This phase transition is similar to the Gross-Witten phase transition but it happens in a moderately large but finite size spin chain.

Finally, using the relations between the Heisenberg spin chains and various gauge theories such as Chern-Simons theories, supersymmetric gauge theories, etc. it would be fruitful to investigate the implications of the obtained results of this study, in the context of gauge theories. 

\section{Acknowledgement}
The authors deeply appreciate Ezad Shojaei, without whom this study would not have been initiated.

\appendix
\section{Dynamics of XX0 spin chain and nonintersecting Brownian motion}
\label{RXN}
This section provides explanations for one of the basic pillars of this study, i.e. the connection between XX0 model and NIBM. This connection is essential for understanding the interpretation of the phase structure of XX0 model. In fact, our results are interpreted in terms of the dynamics of magnons, which is equivalent to the dynamics of vicious walkers in NIBM. In order to describe the relation between the dynamics of XX0 model and NIBM, we follow a mathematical description for the process of measurement in quantum mechanics of spin chains. Roughly speaking, this relation is originated in the probabilistic nature of measurement in quantum mechanics and stochastic nature of NIBM.

\begin{figure}
\begin{center}
\scalebox{0.5}{\includegraphics[angle=0]{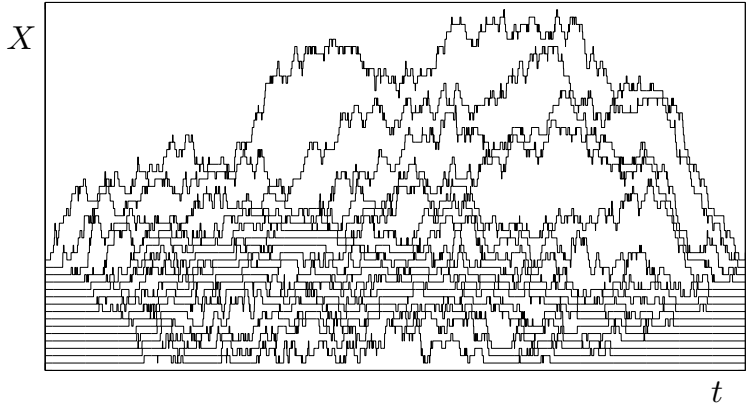}}
\end{center}
\caption{Nonintersecting Brownian motion with fixed initial and final boundary conditions.   
}
\label{nibm1}
\end{figure}

In the context of our study, NIBM is the propagation of $N_{f}$ vicious walkers, (i.e. random walkers such that their paths do not intersect each other) on an one-dimensional lattice with size $N$ and time scale $t$.
Due to the one-dimensional degree of freedom in NIBM,
the random walk of each particle in NIBM is confined between the back and front particles on the lattice. As a result, their trajectories do not cross each other, as shown in Fig.~\ref{nibm1} for a NIBM with reunion boundary conditions.
This dynamics has been investigated and applied extensively in literature, namely in random matrix theory \unscite{nibm1},\unscite{nibm2},\unscite{nibm3},\unscite{nibm4}, two-dimensional Yang-Mills theory \unscite{Forrester},\unscite{nibm6}, three-dimensional Chern-Simons theory \unscite{zahabi}, KPZ growing model \unscite{KPZ1},\unscite{KPZ2}, etc.

To clarify the relationship between dynamics of XX0 model and NIBM, first we describe random motion (diffusion) of a magnon on a finite segment of XX0 spin chain and its relation to a random walk
in the one-dimensional lattice. Then, we consider two and more magnons to illuminate the nature of nonintersecting motion of magnons in XX0 model and random walkers in NIBM. Finally, we conclude that the dynamics of magnons in XX0 model is equivalent to that of random walkers in NIBM.

\textit{Dynamics of magnons in XX0 spin chain}.
Let us review rudiments of XX0 spin chain. The spin operators act on the space $\bigotimes_{k=0}^{M}\mathds{C}^2$ spanned by $\bigotimes_{k=0}^{M}\ket{s_k}$ where $\ket{s_k}$ can be either plus spin, $\ket{\uparrow}= \left( \begin{array}{ccc}
1  \\
0  \\
\end{array} \right)$,
or minus spin,
$\ket{\downarrow}= \left( \begin{array}{ccc}
0 \\
1  \\
\end{array} \right)$. For example, the actions of the annihilation and creation operators on the up-spin states are ${\sigma}_{m}^{-}\ket{\uparrow}_{m}=\ket{\downarrow}_{m}$,
and
${\sigma}_{m}^{+}\ket{\uparrow}_{m}=0$, respectively.

Let us consider the Hamiltonian of XX0 model, Eq. \ref{H_XX0 eq1}. By definition, the action of operator ${\sigma }_{n}^{+}{\sigma }_{n\pm1}^{-}$ on two-spin state 
is to exchange the direction of spins when the spins are in opposite directions, otherwise, the action of operator give the null state,
\begin{equation}
\label{Copling term}
{\sigma }_{n}^{+}{\sigma }_{n+1}^{-}\ket{\underset{n}\downarrow\downarrow}=0,\quad{\sigma }_{n}^{+}{\sigma }_{n+1}^{-}\ket{\underset{n}\uparrow\uparrow}=0,\quad{\sigma }_{n}^{+}{\sigma }_{n+1}^{-}\ket{\underset{n}\downarrow\uparrow}=\ket{\underset{n}\uparrow\downarrow},\quad {\sigma }_{n}^{+}{\sigma }_{n-1}^{-}\ket{\uparrow\underset{n}\downarrow}=\ket{\downarrow\underset{n}\uparrow}.
\end{equation}
Now, consider a state with a magnon in the $n$'th site of a ferromagnetic vacuum. The action of Hamiltonian Eq.~\ref{H_XX0 eq1} on this state leads to the entanglement of two different states as follow
\begin{equation}
 \label{action}
 ({\sigma }_{n}^{+}{\sigma }_{n+1}^{-}+{\sigma }_{n}^{+}{\sigma }_{n-1}^{-})\ket{\uparrow,...,\uparrow\underset{n}\downarrow\uparrow,...,\uparrow}\xrightarrow{measurement} 
  \begin{cases}
   \ket{\uparrow,...,\uparrow\underset{n}\uparrow\downarrow,...,\uparrow} & \quad 50\% \\
   \\
   \ket{\uparrow,...,\downarrow\underset{n}\uparrow\uparrow,...,\uparrow} & \quad 50\% \\
  \end{cases}.
\end{equation}
Hence, in the measurement process, the system has to pick up one of these states with the equal probability. Therefore, under the successive measurement, the position of a magnon moves through the ferromagnetic vacuum in spin chain exactly similar to a simple random walk. So far, we have seen that the diffusion of a magnon in XX0 spin chain is equivalent to an one-dimensional random walk. Now, we take one step forward and put two magnons in the ferromagnetic vacuum.
In this situation, each magnon separately moves like a random walker except when the positions of magnons arrive to the adjacent sites.
In this case, the action of ${\sigma }_{n}^{+}{\sigma }_{n\pm1}^{-}$ on this state is
\begin{equation}
\label{Copling term2}
{\sigma }_{n}^{+}{\sigma }_{n+1}^{-}\ket{\uparrow\underset{n}\downarrow\downarrow\uparrow}=0,\quad
{\sigma }_{n}^{+}{\sigma }_{n-1}^{-}\ket{\uparrow\downarrow\underset{n}\downarrow\uparrow}=0,\quad
{\sigma }_{n}^{+}{\sigma }_{n+1}^{-}\ket{\uparrow\downarrow\underset{n}\downarrow\uparrow}=\ket{\uparrow\downarrow\underset{n}\uparrow\downarrow},\quad
{\sigma }_{n}^{+}{\sigma }_{n-1}^{-}\ket{\uparrow\underset{n}\downarrow\downarrow\uparrow}=\ket{\downarrow\underset{n}\uparrow\downarrow\uparrow}.
\end{equation}
As a result
\begin{equation}
 \begin{split}
  \label{action2}
  ({\sigma }_{n}^{+}{\sigma }_{n+1}^{-}+{\sigma }_{n}^{+}{\sigma }_{n-1}^{-})\ket{\uparrow,...,\uparrow\downarrow\underset{n}\downarrow\uparrow,...,\uparrow}\xrightarrow{measurement}\ket{\uparrow,...,\uparrow\downarrow\underset{n}\uparrow\downarrow,...,\uparrow},  \\
  ({\sigma }_{n}^{+}{\sigma }_{n+1}^{-}+{\sigma }_{n}^{+}{\sigma }_{n-1}^{-})\ket{\uparrow,...,\uparrow\underset{n}\downarrow\downarrow\uparrow,...,\uparrow}\xrightarrow{measurement}\ket{\uparrow,...,\downarrow\underset{n}\uparrow\downarrow\uparrow,...,\uparrow}. 
 \end{split}
\end{equation}
This means, as soon as two magnons arrive to each other, they act to get away from the contact point. In other words, corresponding random walkers do not cross their trajectories. In the case of more than two magnons, the motion of all magnons is similar to one-dimensional random walk with a condition, i.e. their path do not intersect each other. This exactly defines the vicious (nonintersecting) random walk
and illuminates the nonintersecting character of magnons diffusion in the vacuum.
To summarize, the vicious walkers in one-dimensional lattice in NIBM are equivalent to diffusing magnons in ferromagnetic vacuum of XX0 spin chain.

\section{Tracy-Widom distribution}
\label{TWi}
The Tracy-Widom distribution function can be defined as
\begin{eqnarray}
\label{TWidd}
F(s)=\exp\left(-\int_{s}^{\infty}(x-s)q^{2}(x)dx\right),
\end{eqnarray}
where $q(s)$ is the solution of Painlev\'{e} equation
\begin{eqnarray}
\label{q}
q^{''}(s)=sq(s)+2q(s)^{3}\hspace{.5cm} \textit{with}\hspace{.5cm} q(s)\sim -Ai(s)\hspace{.5cm} \textit{as}\hspace{.5cm} x\rightarrow\infty,
\end{eqnarray}
where $Ai(s)$ is the Airy function.
This distribution function is introduced and elaborated in detail in \unscite{Tracy},\unscite{Tracy2}.

A precise asymptotic analysis of the Tracy-Widom function is obtained in \unscite{Baik3}, 
\begin{equation}
 \label{asym TW}
 F(x) = 
  \begin{cases}
   1-\frac{e^{-\frac{4}{3}x^{\frac{3}{2}}}}{32\pi x^{\frac{3}{2}}}(1-\frac{35}{24x^{\frac{3}{2}}})+\mathcal{O}(x^{-3})                       & \quad x \rightarrow \infty \\
   \\
   2^{\frac{1}{42}}e^{\zeta(-1)}\frac{e^{-\frac{1}{12}|x|^{3}}}{|x|^{\frac{1}{8}}} (1-\frac{3}{2^{6}|x|^{-3}})+\mathcal{O}(x^{-6})      & \quad x \rightarrow -\infty \\
  \end{cases},
\end{equation}
where $\zeta$ is the Riemann zeta function. This asymptotic analysis of Tracy-Widom distribution implies that
\begin{equation}
\label{ATW}
F(x) =
  \begin{cases}
   1-\mathcal{O}(e^{-x^{3/2}}) & \quad x \rightarrow \infty \\
   \\
   \mathcal{O}(e^{-|x|^{3}}) & \quad x \rightarrow -\infty \\
  \end{cases}.
\end{equation}

The Tracy-Widom distribution function appears in various topics such as random matrix theory, vicious walkers \unscite{Majumdar}, gauge theories \unscite{zahabi},\unscite{Forrester}, etc. In random matrix theory this distribution function appears as the limiting distribution, $\rho(\alpha)$, for the largest eigenvalue, $\alpha$, of $N_f\times N_f$ Hermitian matrices in the Gaussian Unitary Ensemble (GUE). More precisely,
$\frac{\alpha-\sqrt{2N_f}}{(\sqrt{2})^{-1}N_f^{-\frac{1}{6}}}
\xrightarrow{\mathcal{D}} \mathcal{TW}_2$.

\begin{figure}
\begin{center}
\scalebox{0.5}{\includegraphics[angle=0]{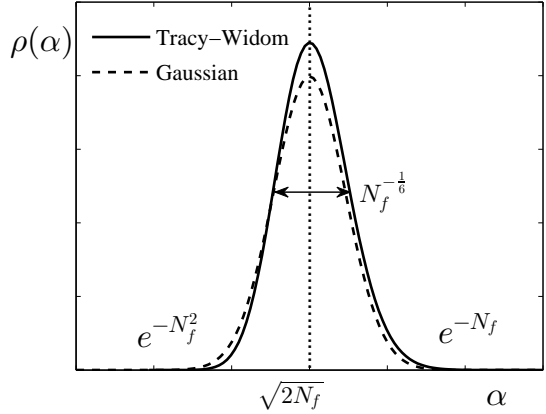}}
\end{center}
\caption{Tracy-Widom distribution of the largest eigenvalue in GUE, compared with Gaussian distribution function.
}
\label{TWpf}
\end{figure}
As can be seen in Eq. \ref{asym TW} and also in Fig.~\ref{TWpf}, the two tails of the Tracy-Widom distribution function decay differently and this leads to the phase transition or cross-over phenomena in different models that are governed by this distribution in their asymptotic limit.

\section{Toeplitz and Hankel determinants in nonintersecting Brownian motion}
\label{MIG}
In this section we review the obtained results by Baik and Liu \unscite{Baik}. They studied probability distribution of width of NIBM by using Toeplitz and Hankel determinants. We present an adopted version of these results with the parameters of XX0 model. These results are used in our study to obtain the asymptotic limit of the partition functions, the phase structure of XX0 model and its interpretation.
\subsection{Toeplitz determinants and nonintersecting simple symmetric random walks}
The continuous and discrete Toeplitz determinant are defined as
\begin{eqnarray}
\label{Teoplitz}
D_{N_{f}}({\textit{f}})=det\begin{bmatrix}\int_{|z|=1}z^{-j+l}f(z)\frac{dz}{2\pi iz}\end{bmatrix}_{j,l=0}^{N_{f}-1},\quad D_{N_{f}}^{|d|}(f,|d|)=det\begin{bmatrix}\frac{1}{|d|} \sum_{z \in d} z^{-j+l}f(z)\end{bmatrix}_{j,l=0}^{N_{f}-1},
\end{eqnarray}
where $d$ is a finite domain with the size $|d|$ and $f(z)$ is the weight function. By Heine-Szeg\"{o} identity \unscite{Heine}, the Toeplitz determinants are equivalent to matrix integrals and this fact is used in our study for representing partition function of strongly coupled XX0 model by Toeplitz determinant including Gross-Witten weight function. 

Let us review some necessary facts and results in nonintersecting continuous-time symmetric simple random walkers $X(t'):=(X_0(t'), X_1(t'), ..., X_{N_f-1}(t'))$. The distance of these nonintersecting walkers away from origin are defined as $X_{i}(t), i=0,...,N_f-1$ and they are subject to boundary conditions, $X(0)=X(T)=(0, 1, ..., N_f-1)$. The nonintersecting character of these walkers imply the condition;
$X_{0}(t')<X_{1}(t')<...<X_{N_{f}-1}(t')$ for all $t' \in [0,t]$. The maximum distance between first and last walker is defined as a width $W_{N_{f}}(t)=sup_{t' \in [0,t]}(X_{N_{f}-1}(t')-X_{1}(t'))$. It has been proved in \unscite{Baik} that in a domain with size $N$ the conditional probability on $W_{N_{f}}(t)<N$ can be expressed in terms of Toeplitz determinants, as
\begin{eqnarray}
\label{con prob1}
\mathbb{P}(W_{N_{f}}(t)<N) = \oint_{|s|=1} \frac{D_{N_{f}}^{|d_{s}|}(f,|d_{s}|)}{D_{N_{f}}(f)} \frac{ds}{2\pi i s}, \hspace{.5cm} f(z)=e^{\frac{t}{2}(z+z^{-1})},
\end{eqnarray}
where $d_{s}=\{z\in \mathds{C} |z^{N}=s\}$ and $|d_s|=N$. The event that $W_{N_f}<N$ confines the random walkers into a chamber; $0\leq X_{0}<X_{1}<...<X_{N_{f}-1}<X_0+N$. Furthermore, the asymptotic limit of the conditional probability is obtained as
\begin{eqnarray}
\label{con prob2}
\lim_{min(N_{f},t)\to\infty}\mathbb{P}\left(\frac{W_{N_{f}}(t)-\mu (N_{f},t)}{\sigma (N_{f},t)}\leqslant x\right)=F(x),
\end{eqnarray}
where $F(x)$ is the Tracy-Widom distribution Function \unscite{Tracy} and 
\begin{equation}
 \label{mu-sigma}
 \mu  \coloneqq
  \begin{cases}
   N_{f}+t & \quad N_{f} \geq t\\
   2\sqrt{N_{f}t} & \quad N_{f} < t\\
  \end{cases},
   \quad
 \sigma  \coloneqq
  \begin{cases}
   2^{-\frac{1}{3}}t^{\frac{1}{3}} &  \quad N_{f} \geq t\\
   2^{-\frac{2}{3}}t^{\frac{1}{3}}(\sqrt{\frac{N_{f}}{t}}+\sqrt{\frac{t}{N_{f}}})^{\frac{1}{3}} & \quad N_{f} < t\\
  \end{cases}.
\end{equation}
We used the above obtained results in the case $s=1$, in a two-fold way, first for deriving the asymptotic limit of the partition function of strongly coupled finite XX0 model and second for interpreting the obtained results in terms of the propagation of magnons in the spin chain. 

\subsection{Hankel determinants and nonintersecting Brownian bridges}
The continuous and discrete Hankle Determinant are defined as
\begin{eqnarray}
\label{Hankel}
\mathcal{H}_{N_{f}}({\textit{f}})=det\begin{bmatrix}\int_{\mathbb{R}}x^{j+k}f(x)dz\end{bmatrix}_{j,k=0}^{N_{f}},\quad H_{N_{f}}({\textit{f},d})=det\begin{bmatrix}\sum_{x \in d} x^{j+k}f(x)\end{bmatrix}_{j,k=0}^{N_{f}},
\end{eqnarray}
where $d$ is a discrete subset of $\mathbb{R}$. Heine-Szeg\"{o} identity \unscite{Heine}, identifies the discrete and continuous Hankel determinants with discrete and continuous matrix integrals. We used this fact, to study the partition function of weakly coupled XX0 model, represented as Gaussian matrix integral, by Hankel determinant. 

Let us review the definitions and results for probability distribution of width in nonintersecting Brownian bridges in terms of the Hankel determinants \unscite{Baik}. The distance of random walkers from the origin is defined as $X_{i}(t), i=1,...,N_f$, with boundary conditions $X_i(0)=X_i(1)=0$ for all $i=1, ..., N_f$, and they are conditioned that $X_{1}(t')<X_{2}(t')<...<X_{N_{f}}(t')$ for all $t' \in (0,1)$. The width of this process is defined as $W_{N_{f}} \coloneqq \underset{0\leq t' \leq 1}{\mathrm{sup}}(X_{N_{f}}(t')-X_{1}(t'))$.
The conditional probability distribution of the width of nonintersecting Brownian bridges is obtained in the terms of the Hankel determinants, \unscite{Karlin},\unscite{Hobson}, as
\begin{eqnarray}
\label{Hankel con prob1}
\mathbb{P}(W_{N_{f}}<N)=\frac{(\frac{\sqrt{2}\pi}{N\sqrt{N_{f}}})^{N_{f}}}{\mathcal{H}_{N_{f}}(f)} \int_{0}^{1} H_{N_{f}}(f,|d_{s}|) ds, \hspace{.5cm} f(x)=e^{-N_fx^2},
\end{eqnarray}
where the discrete domain is defined by $d_{s}\coloneqq \{ \frac{\sqrt{2}\pi}{N\sqrt{N_{f}}}(m-s)|m \in \mathbb{Z} \}$ and the size of the domain is $|d_s|=N\sqrt{N_f}$. 
The asymptotic limit of the conditional probability is obtained in \unscite{Baik}, as
\begin{eqnarray}
\label{Hankel con prob2}
\lim_{N_{f}\to\infty}\mathbb{P}\left((W_{N_{f}}-2\sqrt{N_{f}})2^{2/3}N_{f}^{1/6}\leqslant x\right)=F(x),
\end{eqnarray}
where $F(x)$ is the Tracy-Widom distribution function \unscite{Tracy},\unscite{Baik}.
In this study, we apply above obtained results in case $s=0$, for studying the asymptotic limit of the weakly coupled XX0 model and and its interpretation in terms of behaviors of the diffusing magnons in the spin chain. 

\section{Matrix integral representation of XX0 correlation functions}
\label{PfCf}
In this section we will review the matrix integral representation for the correlation functions of XX0 spin chain. A general correlation function of XX0 model is defined as
\begin{equation}
\label{C}
C_{j_{1},...,j_{N_{f}};l_{1},...,l_{N_{f}}}\coloneqq\bra{\Uparrow}\sigma^{+}_{j_{1}}...\sigma^{+}_{j_{N_{f}}}e^{-tH_{XX0}}\sigma^{-}_{l_{1}}...\sigma^{-}_{l_{N_{f}}}\ket{\Uparrow},
\end{equation}
where $\ket{\Uparrow}$ is a ferromagnetic state of size $N$, $H_{XX0}$ is the Hamiltonian and $\sigma^{\pm}$ are the spin operators act on $l_{1},...,l_{N_{f}}$ ($j_{1},...,j_{N_{f}}$) positions in the initial (final) ferromagnetic state and flip $N_f$ spins in these positions.  
In the limit of large size, $N\gg 1$, and for $\Delta=1$, the matrix integral representation of the general correlation function, $C_{j_{1},...,j_{N_{f}};l_{1},...,l_{N_{f}}}$, is obtained in \unscite{Bogoliubov}, as
\begin{eqnarray}
\label{CM}
\frac{1}{N_f!}\bigg( \frac{1}{2\pi} \bigg)^{N_{f}}\int^{\pi}_{-\pi}d\theta_{1}...
                                            \int^{\pi}_{-\pi}d\theta_{N_{f}}&&
                               S_{\lambda}(e^{i\theta_{1}},e^{i\theta_{2}},...,e^{i\theta_{N_{f}}})
                                S_{\lambda'}(e^{-i\theta_{1}},e^{-i\theta_{2}},...,e^{-i\theta_{N_{f}}})\nonumber\\
                                            &&e^{t\sum_{m=1}^{N_{f}}cos_{}\theta_{m}}\prod_{1\leq j < k \leq N_{f}}|e^{i\theta_{j}}-e^{i\theta_{k}}|^{2},
\end{eqnarray}
where $\theta_{i}$'s are eigenvalues of the random matrix, $S_\lambda$ is the symmetric Schur function of strict partition $\lambda=(N\geq\lambda_1>\lambda_2>...>\lambda_{N_f}\geq 0)$ with $\lambda_i=j_i-N_f+i$ and $\lambda'_i=l_i-N_f+i$ ($i=0, 1, ..., N_f-1$). The matrix integral (\ref{CM}) contains symmetric Schur functions \unscite{macdonald}, and these functions in the matrix integral play the role of spin operators in XX0 model and in fact they represent and fix the locations of the flipped spins, in the correlation function.
We can choose the locations of spin operators as in the partition function of XX0 model, Eq.~\ref{Delta eq7}, then we observe that $\lambda_i=\lambda'_i=0$ and thus the Schur functions in expression (\ref{CM}) become one. Therefore, the correlation function (\ref{C}) and (\ref{CM}) with this choice, which is by definition the partition function of the model, reduces to a matrix model with Gross-Witten potential as following
\begin{eqnarray}
\label{CMI}
\Z_{XX0}&=&\bra{\overbrace{{\uparrow},...,{\uparrow},\underbrace{{\downarrow},...,{\downarrow},{\downarrow}}_{N_f}}^{N}}e^{-t{\hat{H}}_{XX0}}\ket{\overbrace{\underbrace{{\downarrow},{\downarrow},...,{\downarrow}}_{N_f},{\uparrow},...,{\uparrow}}^{N}} \nonumber \\
  &=&\frac{1}{N_f!}\bigg( \frac{1}{2\pi} \bigg)^{N_{f}}\int^{\pi}_{-\pi}d\theta_{1}...
                                            \int^{\pi}_{-\pi}d\theta_{N_{f}}e^{t\sum_{m=1}^{N_{f}}cos_{}\theta_{m}}
                                            \prod_{1\leq j < k \leq N_{f}}|e^{i\theta_{j}}-e^{i\theta_{k}}|^{2}.
\end{eqnarray}

\label{s7}
\mybibitem{TW1}{Tracy, Craig A., and Harold Widom. "Nonintersecting brownian excursions." The Annals of Applied Probability 17, no. 3 (2007): 953-979.}
\mybibitem{Schehr1}{Schehr, Grégory, Satya N. Majumdar, Alain Comtet, and Julien Randon-Furling. "Exact distribution of the maximal height of p vicious walkers." Physical review letters 101, no. 15 (2008): 150601.}
\mybibitem{Kobayashi}{Kobayashi, Naoki, Minami Izumi, and Makoto Katori. "Maximum distributions of bridges of noncolliding Brownian paths." Physical Review E 78, no. 5 (2008): 051102.}
\mybibitem{Wadia}{Wadia, Spenta R. "N=∞ phase transition in a class of exactly soluble model lattice gauge theories." Physics Letters B 93, no. 4 (1980): 403-410.}
\mybibitem{Schehr}{Majumdar, Satya N., and Grégory Schehr. "Top eigenvalue of a random matrix: large deviations and third order phase transition." Journal of Statistical Mechanics: Theory and Experiment 2014, no. 1 (2014): P01012.}
\mybibitem{kitanine}{Kitanine, N., J. M. Maillet, N. A. Slavnov, and V. Terras. "Correlation functions of the XXZ spin-12 Heisenberg chain at the free fermion point from their multiple integral representations." Nuclear Physics B 642, no. 3 (2002): 433-455.}
\mybibitem{colomo1}{Colomo, F., A. G. Izergin, V. E. Korepin, and V. Tognetti. "Correlators in the Heisenberg XXO chain as Fredholm determinants." Physics Letters A 169, no. 4 (1992): 243-247.}
\mybibitem{colomo2}{Colomo, Filippo, Anatolii Georgievich Izergin, Vladimir Evgen'evich Korepin, and Valerio Tognetti. "Temperature correlation functions in the XX0 Heisenberg chain. I." Theoretical and Mathematical Physics 94, no. 1 (1993): 11-38.}
\mybibitem{colomo3}{Colomo, F., A. G. Izergin, and V. Tognetti. "Correlation functions in the XXO Heisenberg chain and their relations with spectral shapes." Journal of Physics A: Mathematical and General 30, no. 2 (1997): 361.}
\mybibitem{Forrester}{Forrester, Peter J., Satya N. Majumdar, and Grégory Schehr. "Non-intersecting Brownian walkers and Yang–Mills theory on the sphere." Nuclear Physics B 844, no. 3 (2011): 500-526.}
\mybibitem{macmahon}{MacMahon, Percy Alexander. Combinatory analysis. Vol. 2. Courier Corporation, 1916.}
\mybibitem{colomo}{Colomo, F., and A. G. Pronko. "Third-order phase transition in random tilings." Physical Review E 88, no. 4 (2013): 042125.}
\mybibitem{baxter}{Baxter, Rodney J. Exactly solved models in statistical mechanics. Courier Corporation, 2007.}
\mybibitem{schulz}{Schulz, H. J. "Mesoscopic quantum physics." Les Houches Proceedings, edited by E. Akkermans et al.(Elsevier) (1995).}
\mybibitem{faddeev}{Faddeev, L. D. "How algebraic Bethe ansatz works for integrable model." arXiv preprint hep-th/9605187 (1996).}
\mybibitem{sachdev}{Sachdev, Subir. Quantum phase transitions. John Wiley - Sons, Ltd, 2007.}
\mybibitem{eisert}{Eisert, Jens, Marcus Cramer, and Martin B. Plenio. "Colloquium: Area laws for the entanglement entropy." Reviews of Modern Physics 82, no. 1 (2010): 277.}
\mybibitem{minahan}{Minahan, Joseph A. "Review of AdS/CFT Integrability, Chapter I. 1: Spin Chains in {\ mathcal {N}= 4} Super Yang-Mills." Letters in Mathematical Physics 99, no. 1-3 (2012): 33-58.}
\mybibitem{nekrasov1}{Nekrasov, Nikita, and Samson Shatashvili. "Quantum integrability and supersymmetric vacua." Progress of Theoretical Physics Supplement 177 (2009): 105-119.}
\mybibitem{nekrasov2}{Nekrasov, Nikita A., and Samson L. Shatashvili. "Supersymmetric vacua and Bethe ansatz." arXiv preprint arXiv:0901.4744 (2009).}
\mybibitem{bogolyobovRev}{Bogolyubov, Nikolai Mikhailovich, and Cyril Leonidovich Malyshev. "Integrable models and combinatorics." Russian Mathematical Surveys 70, no. 5 (2015): 789.}
\mybibitem{borodin}{Borodin, Alexei, and Vadim Gorin. "Lectures on integrable probability." (2012).}
\mybibitem{di}{Di Francesco, Philippe. "Integrable combinatorics." In XVIIth International Congress on Mathematical Physics, World Sci. Publ., Hackensack, NJ, pp. 29-51. 2014.}
\mybibitem{blower}{Blower, Gordon. Random matrices: high dimensional phenomena. Vol. 367. Cambridge University Press, 2009. Harvar}
\mybibitem{forrester}{Forrester, Peter J. Log-gases and random matrices (LMS-34). Princeton University Press, 2010.}
\mybibitem{fisher}{Fisher, Michael E. "Walks, walls, wetting, and melting." Journal of Statistical Physics 34, no. 5-6 (1984): 667-729.}
\mybibitem{krattenthaler}{Krattenthaler, Christian. The major counting of nonintersecting lattice path and generating functions for tableaux. Vol. 115. Providence (RI): Amer. math. soc., 1995.}
\mybibitem{baik}{Baik, Jinho, Percy Deift, and Toufic Suidan. Combinatorics and Random Matrix Theory. Vol. 172. American Mathematical Soc., 2016.}
\mybibitem{okounkov}{Okounkov, Andrei, Nikolai Reshetikhin, and Cumrun Vafa. "Quantum Calabi-Yau and classical crystals." In The unity of mathematics, pp. 597-618. Birkhäuser Boston, 2006.}
\mybibitem{bressoud}{Bressoud, David M. Proofs and Confirmations: The Story of the Alternating-Sign Matrix Conjecture. Cambridge University Press, 1999.}

\mybibitem{johansson} {Johansson, Kurt. "The longest increasing subsequence in a random permutation and a unitary random matrix model." Mathematical Research Letters 5 (1998): 63-82.
}
\mybibitem{majumdar}{
Schehr, Grégory, Satya N. Majumdar, Alain Comtet, and Peter J. Forrester. "Reunion probability of N vicious walkers: typical and large fluctuations for large N." Journal of Statistical Physics 150, no. 3 (2013): 491-530.
}
\mybibitem{baik-suidan} {Baik, Jinho, and Toufic M. Suidan. "Random matrix central limit theorems for nonintersecting random walks." The Annals of Probability 35, no. 5 (2007): 1807-1834.
}

\mybibitem{Q PhT}{ Werlang, T., C. Trippe, G. A. P. Ribeiro, and Gustavo Rigolin. "Quantum correlations in spin chains at finite temperatures and quantum phase transitions." Physical review letters 105, no. 9 (2010): 095702.}

\mybibitem{T PhT} {Haldane, F. D. M. "Nonlinear field theory of large-spin Heisenberg antiferromagnets: semiclassically quantized solitons of the one-dimensional easy-axis Néel state." Physical Review Letters 50, no. 15 (1983): 1153.}

\mybibitem{Schollwock}{Mikeska, Hans-Jürgen, Alexei K. Kolezhuk, Ulrich Schollwock, Johannes Richter, Damian JJ Farnell, and Raymod F. Bishop. "Quantum Magnetism." (2004).
Harvard	
}


\mybibitem{KPZ1} {Corwin, Ivan, and Alan Hammond. "Brownian Gibbs property for Airy line ensembles." Inventiones mathematicae 195, no. 2 (2014): 441-508.}

\mybibitem{KPZ2}{Corwin, Ivan. "The Kardar–Parisi–Zhang equation and universality class." Random matrices: Theory and applications 1, no. 01 (2012): 1130001.}

\mybibitem{nibm1}{Baik, Jinho. "Random vicious walks and random matrices." arXiv preprint math/0001022 (2000).}
\mybibitem{nibm2}{Forrester, P. J. "Random walks and random permutations." Journal of Physics A: Mathematical and General 34, no. 31 (2001): L417.}
\mybibitem{nibm3}{Johansson, Kurt. "Discrete polynuclear growth and determinantal processes." Communications in Mathematical Physics 242, no. 1-2 (2003): 277-329.}
\mybibitem{nibm4}{Nagao, Taro. "Dynamical correlations for vicious random walk with a wall." Nuclear Physics B 658, no. 3 (2003): 373-396.}

\mybibitem{nibm5}{Forrester, Peter J., Satya N. Majumdar, and Grégory Schehr. "Non-intersecting Brownian walkers and Yang–Mills theory on the sphere." Nuclear Physics B 844, no. 3 (2011): 500-526.}
\mybibitem{nibm6}{de Haro, Sebastian, and Miguel Tierz. "Brownian motion, Chern–Simons theory, and 2d Yang–Mills." Physics Letters B 601, no. 3 (2004): 201-208.}

\mybibitem{Lieb}
{E. Lieb, T. Schultz, D. Mattis, Ann. Phys. (NY) 16 (1961) 407.}


\mybibitem{macdonald}{Macdonald, Ian Grant. Symmetric functions and Hall polynomials. Oxford university press, 1998.}
\mybibitem {Bressoud} {Bressoud, David M. Proofs and Confirmations: The Story of the Alternating-Sign Matrix Conjecture. Cambridge University Press, 1999.}
\mybibitem {Krattenthaler}{
Krattenthaler, Christian, Anthony J. Guttmann, and Xavier G. Viennot. "Vicious walkers, friendly walkers and Young tableaux: II. With a wall." Journal of Physics A: Mathematical and General 33, no. 48 (2000): 8835.}
\mybibitem{Bogoliubov}{
Bogoliubov, N. M. "XX0 Heisenberg chain and random walks." Journal of Mathematical Sciences 138, no. 3 (2006): 5636-5643.}
\mybibitem{Bogoliubov1}{
Bogoliubov, N. M. "Integrable models for vicious and friendly walkers." Journal of Mathematical Sciences 143, no. 1 (2007): 2729-2737.}


\mybibitem{Vaidya}{
Vaidya, Hemant G., and C. A. Tracy. "One particle reduced density matrix of impenetrable bosons in one dimension at zero temperature." Journal of Mathematical Physics 20, no. 11 (1979): 2291-2312.}
\mybibitem{McCoy}{
McCoy, Barry M., Jacques HH Perk, and Robert E. Shrock. "Time-dependent correlation functions of the transverse Ising chain at the critical magnetic field." Nuclear Physics B 220, no. 1 (1983): 35-47.}
\mybibitem{McCoy1}{
McCoy, Barry M., Jacques HH Perk, and Robert E. Shrock. "Correlation functions of the transverse Ising chain at the critical field for large temporal and spatial separations." Nuclear Physics B 220, no. 3 (1983): 269-282.}

\mybibitem{Its}{Izergin, A. G., A. R. Its, V. E. Korepin, and N. A. Slavnov. "Integrable differential equations for temperature correlation functions of the XXO Heisenberg chain." Journal of Mathematical Sciences 80, no. 3 (1996): 1747-1759.}
\mybibitem{Its1}{Its, A. R., A. G. Izergin, and V. E. Korepin. "Temperature correlators of the impenetrable Bose gas as an integrable system." Communications in mathematical physics 129, no. 1 (1990): 205-222.}
\mybibitem{Its2}{Its, A. R., A. G. Izergin, V. E. Korepin, and N. A. Slavnov. "Differential equations for quantum correlation functions." International Journal of Modern Physics B 4, no. 05 (1990): 1003-1037.}
\mybibitem{Its3}{Its, A. R., A. G. Izergin, and V. E. Korepin. "Space correlations in the one-dimensional impenetrable Bose gas at finite temperature." Physica D: Nonlinear Phenomena 53, no. 1 (1991): 187-213.}
\mybibitem{Its4}{Its, A. R., A. G. Izergin, V. E. Korepin, and G. G. Varzugin. "Large time and distance asymptotics of field correlation function of impenetrable bosons at finite temperature." Physica D: Nonlinear Phenomena 54, no. 4 (1992): 351-395.}
\mybibitem{Izergin}{Izergin, A. G., A. R. Its, V. E. Korepin, and N. A. Slavnov. "Integrable differential equations for temperature correlation functions of the XXO Heisenberg chain." Journal of Mathematical Sciences 80, no. 3 (1996): 1747-1759.}

\mybibitem{Lenard0}{Lenard, Andrew. "Momentum Distribution in the Ground State of the One-Dimensional System of Impenetrable Bosons." Journal of Mathematical Physics 5, no. 7 (1964): 930-943.}
\mybibitem{Lenard}{Lenard, A. "One-Dimensional Impenetrable Bosons in Thermal Equilibrium." Journal of Mathematical Physics 7, no. 7 (1966): 1268-1272.}
\mybibitem{Korepin}{Korepin, Vladimir E., and N. A. Slavnov. "The time dependent correlation function of an impenetrable Bose gas as a Fredholm minor. I." Communications in mathematical physics 129, no. 1 (1990): 103-113.}


%
\mybibitem{RM Gravity}{Di Francesco, Philippe, Paul Ginsparg, and Jean Zinn-Justin. "2D gravity and random matrices." Physics Reports 254, no. 1 (1995): 1-133.}
\mybibitem{RM QCD}{Zafeiropoulos, Savvas. "Random Matrix Theories for Lattice QCD Dirac Operators." PhD diss., Stony Brook University, 2013. Harvard}
\mybibitem{Rm network}{Tulino, Antonia M., and Sergio Verdú. Random matrix theory and wireless communications. Vol. 1. Now Publishers Inc, 2004.}
\mybibitem{RM finance1}{Mantegna, Rosario N., and H. Eugene Stanley. Introduction to econophysics: correlations and complexity in finance. Cambridge university press, 1999.}
\mybibitem{RM finance2}{Plerou, Vasiliki, Parameswaran Gopikrishnan, Bernd Rosenow, Luís A. Nunes Amaral, and H. Eugene Stanley. "Universal and nonuniversal properties of cross correlations in financial time series." Physical Review Letters 83, no. 7 (1999): 1471.}
\mybibitem{RM finance3}{Plerou, Vasiliki, Parameswaran Gopikrishnan, Bernd Rosenow, Luis A. Nunes Amaral, Thomas Guhr, and H. Eugene Stanley. "Random matrix approach to cross correlations in financial data." Physical Review E 65, no. 6 (2002): 066126.}
\mybibitem{RM finance4}{Podobnik, Boris, Duan Wang, Davor Horvatic, Ivo Grosse, and H. Eugene Stanley. "Time-lag cross-correlations in collective phenomena." EPL (Europhysics Letters) 90, no. 6 (2010): 68001.}
\mybibitem{RM finance5}{Plerou, Vasiliki, P. Gopikrishnan, Bernd Rosenow, LA Nunes Amaral, and H. Eugene Stanley. "A random matrix theory approach to financial cross-correlations." Physica A: Statistical Mechanics and its Applications 287, no. 3 (2000): 374-382.}
\mybibitem{RM nuclear}{Dyson, Freeman J. "A Brownian-motion model for the eigenvalues of a random matrix." Journal of Mathematical Physics 3, no. 6 (1962): 1191-1198.}
\mybibitem{Schur}{Macdonald, I. G. "Symmetric functions and Hall polynomials Oxford Univ." Press, New York (1995).}

\mybibitem{Heine}{Novak, J. O. N. A. T. H. A. N. "Topics in combinatorics and random matrix theory." (2009).}


\mybibitem{Baik2}{Baik, Jinho, Percy Deift, and Kurt Johansson. "On the distribution of the length of the longest increasing subsequence of random permutations." Journal of the American Mathematical Society 12, no. 4 (1999): 1119-1178.}
\mybibitem{Baik3}{Baik, Jinho, Robert Buckingham, and Jeffery DiFranco. "Asymptotics of Tracy-Widom distributions and the total integral of a Painlevé II function." Communications in Mathematical Physics 280, no. 2 (2008): 463-497.}
\mybibitem{Tracy}{Tracy, Craig A., and Harold Widom. "On orthogonal and symplectic matrix ensembles." Communications in Mathematical Physics 177, no. 3 (1996): 727-754.}
\mybibitem{Tracy2}{Tracy, Craig A., and Harold Widom. "Level-spacing distributions and the Airy kernel." Communications in Mathematical Physics 159, no. 1 (1994): 151-174.}

\mybibitem{Gross}{Gross, David J., and Edward Witten. "Possible third-order phase transition in the large-N lattice gauge theory." Physical Review D 21, no. 2 (1980): 446.}

\mybibitem{Bogoliubov2}{Bogoliubov, N., and C. Malyshev. "The correlation functions of the XXZ Heisenberg chain in the case of zero or infinite anisotropy, and random walks of vicious walkers." St. Petersburg Mathematical Journal 22, no. 3 (2011): 359-377.}
\mybibitem{Majumdar}{Schehr, Grégory, Satya N. Majumdar, Alain Comtet, and Peter J. Forrester. "Reunion probability of N vicious walkers: typical and large fluctuations for large N." Journal of Statistical Physics 150, no. 3 (2013): 491-530.}
\mybibitem{zahabi}{Zahabi, Ali. "New phase transitions in Chern Simons matter theory." Nuclear Physics B 903 (2016): 78-103.}
\mybibitem{Mehta org}{Mehta, M. L. Random Matrices (Pure and applied mathematics, v. 142). Edited by Madan Lal Mehta. Elsevier Science Limited, 2004.}

\mybibitem{Mehta}{Mehta, Madan Lal. Random matrices. Vol. 142. Academic press, 2004.}
\mybibitem{Mehta1}{Forrester, Peter, and S. V. E. N. Warnaar. "The importance of the Selberg integral." Bulletin of the American Mathematical Society 45, no. 4 (2008): 489-534.}
\mybibitem{Karlin}{Karlin, Samuel, and James McGregor. "Coincidence probabilities." Pacific J. Math 9, no. 4 (1959): 1141-1164.}

\mybibitem{Hobson}{Hobson, David G., and Wendelin Werner. "Non-colliding Brownian motions on the circle." Bulletin of the London Mathematical Society 28, no. 6 (1996): 643-650.}
\mybibitem{Douglas}{Douglas, Michael R., and Vladimir A. Kazakov. "Large N phase transition in continuum QCD 2." Physics Letters B 319, no. 1 (1993): 219-230.}

\mybibitem{Baik}{Baik, Jinho, and Zhipeng Liu. "Discrete Toeplitz/Hankel determinants and the width of nonintersecting processes." International Mathematics Research Notices (2013): rnt143.}
\mybibitem{perez}{Pérez-García, David, and Miguel Tierz. "Mapping between the Heisenberg XX spin chain and low-energy QCD." Physical Review X 4, no. 2 (2014): 021050.}
\mybibitem{Korepin}{Jin, B-Q., and Vladimir E. Korepin. "Quantum spin chain, Toeplitz determinants and the Fisher—Hartwig conjecture." Journal of statistical physics 116, no. 1-4 (2004): 79-95.}

\end{document}